\documentclass[twocolumn,tighten]{aastex631}
\usepackage{CJKutf8}
\usepackage{amsmath}
\usepackage{enumitem}
\usepackage{booktabs,pbox}
\usepackage{bm}
\usepackage{colortbl,multirow}

\graphicspath{{./figures/}{./spectra/}}

\newcommand{\sbsc}[1]{_\mathrm{#1}}

\newcommand{\HII}{\ion{H}{2}}

\newcommand{\CO}[2]{\mbox{$\mathrm{CO}\,(#1\text{--}#2)$}}
\newcommand{\HCN}[2]{\mbox{$\mathrm{HCN}\,(#1\text{--}#2)$}}
\newcommand{\HCOp}[2]{\mbox{$\mathrm{HCO^+}\,(#1\text{--}#2)$}}
\newcommand{\CS}[2]{\mbox{$\mathrm{CS}\,(#1\text{--}#2)$}}

\newcommand{\uS}{\mbox{$\rm mJy$}}
\newcommand{\uIco}{\mbox{$\rm K~km~s^{-1}$}}
\newcommand{\uLco}{\mbox{$\rm K~km~s^{-1}~pc^2$}}
\newcommand{\ualphaCO}{\mbox{$\rm M_\odot~pc^{-2}\ (K~km~s^{-1})^{-1}$}}
\newcommand{\uV}{\mbox{$\rm km~s^{-1}$}}
\newcommand{\uM}{\mbox{$\rm M_\odot$}}
\newcommand{\uSig}{\mbox{$\rm M_\odot~pc^{-2}$}}
\newcommand{\urho}{\mbox{$\rm M_\odot~pc^{-3}$}}
\newcommand{\uSFR}{\mbox{$\rm M_\odot~yr^{-1}$}}
\newcommand{\uSigSFR}{\mbox{$\rm M_\odot~yr^{-1}~kpc^{-2}$}}

\accepted{April 10, 2024}

\shorttitle{Hidden Gems on a Ring}
\shortauthors{Sun et al.}

\begin{document}

\title{Hidden Gems on a Ring: Infant Massive Clusters and Their Formation Timeline\\Unveiled by ALMA, HST, and JWST in NGC~3351}

\suppressAffiliations

\newcommand{\Princeton}{\affiliation{Department of Astrophysical Sciences, Princeton University, 4 Ivy Lane, Princeton, NJ 08544, USA}}

\newcommand{\McMaster}{\affiliation{Department of Physics and Astronomy, McMaster University, 1280 Main Street West, Hamilton, ON L8S 4M1, Canada}}

\newcommand{\CITA}{\affiliation{Canadian Institute for Theoretical Astrophysics (CITA), University of Toronto, 60 St George Street, Toronto, ON M5S 3H8, Canada}}

\newcommand{\OSU}{\affiliation{Department of Astronomy, The Ohio State University, 140 West 18th Avenue, Columbus, OH 43210, USA}}

\newcommand{\CCAPP}{\affiliation{Center for Cosmology and Astroparticle Physics (CCAPP), 191 West Woodruff Avenue, Columbus, OH 43210, USA}}

\newcommand{\Alberta}{\affiliation{Department of Physics, University of Alberta, Edmonton, AB T6G 2E1, Canada}}

\newcommand{\ANU}{\affiliation{Research School of Astronomy and Astrophysics, Australian National University, Canberra, ACT 2611, Australia}}

\newcommand{\Arizona}{\affiliation{Steward Observatory, University of Arizona, 933 North Cherry Avenue, Tucson, AZ 85721, USA}}

\newcommand{\ASIAA}{\affiliation{Institute of Astronomy and Astrophysics, Academia Sinica, No. 1, Sec. 4, Roosevelt Road, Taipei 10617, Taiwan}}

\newcommand{\ASTROThreeD}{\affiliation{ARC Centre of Excellence for All Sky Astrophysics in 3 Dimensions (ASTRO 3D), Australia}}

\newcommand{\Bonn}{\affiliation{Argelander-Institut f\"ur Astronomie, Universit\"at Bonn, Auf dem H\"ugel 71, 53121 Bonn, Germany}}

\newcommand{\Carnegie}{\affiliation{Observatories of the Carnegie Institution for Science, 813 Santa Barbara Street, Pasadena, CA 91101, USA}}

\newcommand{\CCA}{\affiliation{Center for Computational Astrophysics, Flatiron Institute, 162 Fifth Avenue, New York, NY 10010, USA}}

\newcommand{\CfA}{\affiliation{Center for Astrophysics $\mid$ Harvard \& Smithsonian, 60 Garden Street, Cambridge, MA 02138, USA}}

\newcommand{\CITEVA}{\affiliation{Centro de Astronomía (CITEVA), Universidad de Antofagasta, Avenida Angamos 601, Antofagasta, Chile}}

\newcommand{\CNRS}{\affiliation{CNRS, IRAP, 9 Av. du Colonel Roche, BP 44346, F-31028 Toulouse cedex 4, France}}

\newcommand{\COOL}{\affiliation{Cosmic Origins Of Life (COOL) Research DAO, coolresearch.io}}

\newcommand{\EPFL}{\affiliation{Institute of Physics, Laboratory for galaxy evolution and spectral modelling, EPFL, Observatoire de Sauverny, Chemin Pegais 51, 1290 Versoix, Switzerland}}

\newcommand{\ESO}{\affiliation{European Southern Observatory, Karl-Schwarzschild Stra{\ss}e 2, D-85748 Garching bei M\"{u}nchen, Germany}}

\newcommand{\Gent}{\affiliation{Sterrenkundig Observatorium, Universiteit Gent, Krijgslaan 281 S9, B-9000 Gent, Belgium}}

\newcommand{\Hawaii}{\affiliation{Institute for Astronomy, University of Hawaii, 2680 Woodlawn Drive, Honolulu, HI 96822, USA}}

\newcommand{\Heidelberg}{\affiliation{Astronomisches Rechen-Institut, Zentrum f\"{u}r Astronomie der Universit\"{a}t Heidelberg, M\"{o}nchhofstra\ss e 12-14, D-69120 Heidelberg, Germany}}

\newcommand{\IAC}{\affiliation{Instituto de Astrof\'isica de Canarias, C/ V\'ia L\'actea s/n, E-38205, La Laguna, Spain}}

\newcommand{\IALP}{\affiliation{Instituto de Astrofísica de La Plata, CONICET--UNLP, Paseo del Bosque S/N, B1900FWA La Plata, Argentina}}

\newcommand{\ICRAR}{\affiliation{International Centre for Radio Astronomy Research, University of Western Australia, 35 Stirling Highway, Crawley, WA 6009, Australia}}

\newcommand{\INAF}{\affiliation{INAF -- Osservatorio Astrofisico di Arcetri, Largo E. Fermi 5, I-50157, Firenze, Italy}}

\newcommand{\IPAC}{\affiliation{Caltech-IPAC, 1200 E. California Blvd. Pasadena, CA 91125, USA}}

\newcommand{\IPARC}{\affiliation{Instituto de F\'{\i}sica de Part\'{\i}culas y del Cosmos IPARCOS, Facultad de Ciencias F\'{\i}sicas, Universidad Complutense de Madrid, E-28040, Spain}}

\newcommand{\IRAM}{\affiliation{Institut de Radioastronomie Millim\'etrique (IRAM), 300 Rue de la Piscine, F-38406 Saint Martin d'H\`eres, France}}

\newcommand{\ITA}{\affiliation{Universit\"{a}t Heidelberg, Zentrum f\"{u}r Astronomie, Institut f\"{u}r Theoretische Astrophysik, Albert-Ueberle-Str 2, D-69120 Heidelberg, Germany}}

\newcommand{\IWR}{\affiliation{Universit\"{a}t Heidelberg, Interdisziplin\"{a}res Zentrum f\"{u}r Wissenschaftliches Rechnen, Im Neuenheimer Feld 205, D-69120 Heidelberg, Germany}}

\newcommand{\JHU}{\affiliation{Department of Physics and Astronomy, The Johns Hopkins University, Baltimore, MD 21218, USA}}

\newcommand{\Kansas}{\affiliation{Department of Physics and Astronomy, University of Kansas, 1251 Wescoe Hall Drive, Lawrence, KS 66045, USA}}

\newcommand{\LAM}{\affiliation{Aix Marseille Univ, CNRS, CNES, LAM (Laboratoire d’Astrophysique de Marseille), Marseille, France}}

\newcommand{\Leiden}{\affiliation{Leiden Observatory, Leiden University, P.O. Box 9513, 2300 RA Leiden, The Netherlands}}

\newcommand{\Liverpool}{\affiliation{Astrophysics Research Institute, Liverpool John Moores University, IC2, Liverpool Science Park, 146 Brownlow Hill, Liverpool L3 5RF, UK}}

\newcommand{\Lyon}{\affiliation{Univ Lyon, Univ Lyon 1, ENS de Lyon, CNRS, Centre de Recherche Astrophysique de Lyon UMR5574, F-69230 Saint-Genis-Laval, France}}

\newcommand{\Maryland}{\affiliation{Department of Astronomy, University of Maryland, 4296 Stadium Drive, College Park, MD 20742, USA}}

\newcommand{\MPE}{\affiliation{Max-Planck-Institut f\"{u}r extraterrestrische Physik, Giessenbachstra{\ss}e 1, D-85748 Garching, Germany}}

\newcommand{\MPIA}{\affiliation{Max-Planck-Institut f\"{u}r Astronomie, K\"{o}nigstuhl 17, D-69117, Heidelberg, Germany}}

\newcommand{\Nagoya}{\affiliation{Department of Physics, Nagoya University, Furo-cho, Chikusa-ku, Nagoya, Aichi 464-8602, Japan}}

\newcommand{\NAOJ}{\affil{National Astronomical Observatory of Japan, 2-21-1 Osawa, Mitaka, Tokyo, 181-8588, Japan}}

\newcommand{\Nichidai}{\affil{Department of Physics, General Studies, College of Engineering, Nihon University, 1 Nakagawara, Tokusada, Tamuramachi, Koriyama, Fukushima, 963-8642, Japan}}

\newcommand{\NRAO}{\affiliation{National Radio Astronomy Observatory, 520 Edgemont Road, Charlottesville, VA 22903, USA}}

\newcommand{\OAN}{\affiliation{Observatorio Astron\'{o}mico Nacional (IGN), C/Alfonso XII, 3, E-28014 Madrid, Spain}}

\newcommand{\ObsParis}{\affiliation{Sorbonne Universit\'{e}, Observatoire de Paris, Universit\'{e} PSL, CNRS, LERMA, F-75014, Paris, France}}

\newcommand{\Oxford}{\affiliation{Sub-department of Astrophysics, Department of Physics, University of Oxford, Keble Road, Oxford OX1 3RH, UK}}

\newcommand{\Rutgers}{\affiliation{Department of Physics and Astronomy, Rutgers, the State University of New Jersey, 136 Frelinghuysen Road, Piscataway, NJ 08854, USA}}

\newcommand{\STScI}{\affiliation{Space Telescope Science Institute, 3700 San Martin Drive, Baltimore, MD 21218, USA}}

\newcommand{\STScIESA}{\affiliation{AURA for the European Space Agency (ESA), Space Telescope Science Institute, 3700 San Martin Drive, Baltimore, MD 21218, USA}}

\newcommand{\Surrey}{\affiliation{Department of Physics, University of Surrey, Guildford GU2 7XH, UK}}

\newcommand{\Sydney}{\affiliation{Sydney Institute for Astronomy, School of Physics A28, The University of Sydney, NSW 2006, Australia}}

\newcommand{\TAPIR}{\affil{TAPIR, California Institute of Technology, Pasadena, CA 91125, USA}}

\newcommand{\Tamkang}{\affiliation{Department of Physics, Tamkang University, No.151, Yingzhuan Rd., Tamsui Dist., New Taipei City 251301, Taiwan}}

\newcommand{\Toulouse}{\affiliation{Universit\'{e} de Toulouse, UPS-OMP, IRAP, F-31028 Toulouse cedex 4, France}}

\newcommand{\Toledo}{\affiliation{University of Toledo, 2801 W. Bancroft St., Mail Stop 111, Toledo, OH 43606, USA}}

\newcommand{\UChile}{\affiliation{Departamento de Astronom\'{i}a, Universidad de Chile, Camino del Observatorio 1515, Las Condes, Santiago, Chile}}

\newcommand{\UCM}{\affiliation{Departamento de F\'{\i}sica de la Tierra y Astrof\'{\i}sica, Universidad Complutense de Madrid, E-28040, Spain}}

\newcommand{\UCSD}{\affiliation{Center for Astrophysics and Space Sciences, Department of Physics,  University of California, San Diego, 9500 Gilman Drive, La Jolla, CA 92093, USA}}

\newcommand{\ULL}{\affiliation{Departamento de Astrof\'isica, Universidad de La Laguna, Av. del Astrof\'isico Francisco S\'anchez s/n, E-38206, La Laguna, Spain}}

\newcommand{\UMass}{\affiliation{University of Massachusetts—Amherst, 710 North Pleasant Street, Amherst, MA 01003, USA}}

\newcommand{\UVa}{\affiliation{University of Virginia, 530 McCormick Road, Charlottesville, VA 22904, USA}}

\newcommand{\Wyoming}{\affiliation{Department of Physics and Astronomy, University of Wyoming, Laramie, WY 82071, USA}}

\newcommand{\Zurich}{\affiliation{Institute for Computational Science, University of Z\"urich, Winterthurerstrasse 190, 8057 Z\"urich, Switzerland}}


\author[0000-0003-0378-4667]{Jiayi~Sun \begin{CJK*}{UTF8}{gbsn}(孙嘉懿)\end{CJK*}}
\altaffiliation{NASA Hubble Fellow}
\Princeton
\McMaster
\CITA

\author[0000-0001-9020-1858]{Hao~He \begin{CJK*}{UTF8}{gbsn}(何浩)\end{CJK*}}
\McMaster

\author[0009-0003-9462-4913]{Kyle~Batschkun}
\McMaster

\author[0000-0003-2508-2586]{Rebecca~C.~Levy}
\altaffiliation{NSF Astronomy and Astrophysics Postdoctoral Fellow}
\Arizona

\author[0000-0001-6527-6954]{Kimberly~Emig}
\altaffiliation{NRAO Jansky Fellow}
\NRAO

\author[0000-0002-0579-6613]{M.~Jimena~Rodr\'iguez}
\STScI
\IALP

\author[0000-0002-8806-6308]{Hamid~Hassani}
\Alberta

\author[0000-0002-2545-1700]{Adam~K.~Leroy}
\OSU
\CCAPP

\author[0000-0002-3933-7677]{Eva~Schinnerer}
\MPIA

\author[0000-0002-0509-9113]{Eve~C.~Ostriker}
\Princeton

\author[0000-0001-5817-0991]{Christine~D.~Wilson}
\McMaster

\author[0000-0002-5480-5686]{Alberto~D.~Bolatto}
\Maryland

\author[0000-0001-8782-1992]{Elisabeth~A.~C.~Mills}
\Kansas

\author[0000-0002-5204-2259]{Erik~Rosolowsky}
\Alberta


\author[0000-0002-2278-9407]{Janice~C.~Lee}
\STScI

\author[0000-0002-5782-9093]{Daniel~A.~Dale}
\Wyoming

\author[0000-0003-3917-6460]{Kirsten~L.~Larson}
\STScIESA

\author[0000-0002-8528-7340]{David~A.~Thilker}
\JHU

\author[0000-0001-7130-2880]{Leonardo~Ubeda}
\STScI

\author[0000-0002-3784-7032]{Bradley~C.~Whitmore}
\STScI

\author[0000-0002-0012-2142]{Thomas~G.~Williams}
\Oxford


\author[0000-0003-0410-4504]{Ashley.~T.~Barnes}
\ESO

\author[0000-0003-0166-9745]{Frank~Bigiel}
\Bonn

\author[0000-0002-5635-5180]{M\'elanie~Chevance}
\ITA
\COOL

\author[0000-0001-6708-1317]{Simon~C.~O.~Glover}
\ITA

\author[0000-0002-3247-5321]{Kathryn~Grasha}
\ANU
\ASTROThreeD

\author[0000-0002-9768-0246]{Brent~Groves}
\ICRAR

\author[0000-0001-9656-7682]{Jonathan~D.~Henshaw}
\Liverpool
\MPIA

\author[0000-0002-4663-6827]{R\'emy~Indebetouw}
\NRAO
\UVa

\author[0000-0002-9165-8080]{Mar\'ia~J.~Jim\'enez-Donaire}
\OAN

\author[0000-0002-0560-3172]{Ralf~S.~Klessen}
\ITA
\IWR

\author[0000-0001-9605-780X]{Eric~W.~Koch}
\CfA

\author[0000-0001-9773-7479]{Daizhong~Liu}
\MPE

\author[0000-0002-4822-3559]{Smita~Mathur}
\OSU
\CCAPP

\author[0000-0002-6118-4048]{Sharon~Meidt}
\Gent

\author[0000-0001-5944-291X]{Shyam~H.~Menon}
\Rutgers
\CCA

\author[0000-0002-3289-8914]
{Justus~Neumann}
\MPIA

\author[0000-0001-5965-3530]{Francesca~Pinna}
\IAC
\ULL
\MPIA

\author[0000-0002-0472-1011]{Miguel~Querejeta}
\OAN

\author[0000-0001-6113-6241]{Mattia~C.~Sormani}
\Surrey

\author[0000-0002-9483-7164]{Robin~G.~Tress}
\EPFL


\correspondingauthor{Jiayi~Sun}
\email{jiayi.sun@princeton.edu}


\begin{abstract}
We study young massive clusters (YMCs) in their embedded ``infant'' phase with $\sim0\farcs1$ ALMA, HST, and JWST observations targeting the central starburst ring in NGC~3351, a nearby Milky Way analog galaxy.
Our new ALMA data reveal 18 bright and compact (sub-)millimeter continuum sources, of which 8 have counterparts in JWST images and only 6 have counterparts in HST images.
Based on the ALMA continuum and molecular line data, as well as ancillary measurements for the HST and JWST counterparts, we identify 14 sources as infant star clusters with high stellar and/or gas masses (${\sim}10^5\;\mathrm{M_\odot}$), small radii (${\lesssim}\,5\;\mathrm{pc}$), large escape velocities ($6{-}10\;\mathrm{km/s}$), and short free-fall times ($0.5{-}1\;\mathrm{Myr}$).
Their multiwavelength properties motivate us to divide them into four categories, likely corresponding to four evolutionary stages from starless clumps to exposed \ion{H}{2} region--cluster complexes.
Leveraging age estimates for HST-identified clusters in the same region, we infer an evolutionary timeline going from $\sim$1--2~Myr before cluster formation as starless clumps, to $\sim$4--6~Myr after as exposed \ion{H}{2} region--cluster complexes.
Finally, we show that the YMCs make up a substantial fraction of recent star formation across the ring, exhibit an non-uniform azimuthal distribution without a very coherent evolutionary trend along the ring, and are capable of driving large-scale gas outflows.
\end{abstract}


\section{Introduction} \label{sec:intro}

Massive star clusters ($M_\star\gtrsim10^4\,\uM$) form in dense, turbulent, high-pressure environments \citep{PortegiesZwart_etal_2010,Longmore_etal_2014,Krumholz_etal_2019}.
This process appears distinct from star formation under lower gas density conditions, with the massive clusters quickly and efficiently convert their gas content into stars before the newly formed stellar population exerts strong feedback \citep[e.g.,][]{Krumholz_etal_2014}.
Understanding the formation of massive clusters can thus provide unique constraints on theoretical models of star formation and stellar feedback.
It can also shed light on how the majority of stellar populations were built up earlier in the cosmic history, when high gas density and turbulence conditions were prevalent \citep[e.g.,][]{Madau_Dickinson_2014}.

To achieve a more complete, observation-grounded understanding of massive cluster formation, it is critical to probe the earliest (``infant'') phase, during which star formation and feedback processes are still ongoing.
This has been challenging for several reasons.
The relevant physical processes happen on a short timescale \citep[$\lesssim$ a few Myr; see e.g.,][]{Krumholz_etal_2014}, making it hard to catch forming YMCs in the infant phase.
Those YMCs that \textit{are} in this phase are still deeply embedded in their natal gas and dust ``cradles'' and thus remain basically invisible at short wavelengths \citep[UV, optical, and sometimes even IR; see][]{Kornei_McCrady_2009,Leroy_etal_2018}.
Last but not least, the physical conditions required for creating YMCs (high gas density, turbulence, and pressure) are rare in the Milky Way and the nearest extragalactic systems \citep{PortegiesZwart_etal_2010}.
Building a sizable sample of infant YMCs thus necessitates obtaining sensitive and resolved observations for these compact objects in more distant galaxies, which can be a tall order even with the latest facilities.

Recent studies based on deep, long-baseline, targeted observations with the Atacama Large Millimeter Array (ALMA) have identified a promising avenue for addressing these challenges.
With its exquisite sensitivity and resolving power in (sub-)millimeter bands, ALMA can detect extinction-free tracers of star formation (via free-free continuum and radio recombination lines) \textit{as well as} the associated gas reservoir (via molecular lines and dust continuum) for YMCs even at extragalactic distances.
Thanks to these unique capabilities, we have already identified and characterized individual forming YMCs or entire YMC populations in our Galaxy \citep[][]{Schmiedeke_etal_2016,Ginsburg_etal_2018} and a handful of nearby galaxies 
(e.g., the Large Magellanic Cloud: \citealt{Nayak_etal_2019};
NGC~5253: \citealt{Turner_etal_2017}; 
NGC~253: \citealt{Leroy_etal_2018,Levy_etal_2021,Mills_etal_2021};
NGC~4945: \citealt{Emig_etal_2020};
Henize~2-10: \citealt{Costa_etal_2021};
and the Antennae: \citealt{Finn_etal_2019,He_etal_2022}).

Building on these pioneering studies, the logical next steps are to connect the infant YMCs identified in the (sub-)millimeter bands with more evolved clusters visible at shorter wavelengths, and to put them into the larger-scale context of the host galaxy.
Doing so requires multiwavelength observations of a sizable YMC population at matched spatial resolution, supported by rich ancillary data for the host galaxy.
It is also important that the host galaxy is at a favorable viewing angle for source localization and cross-wavelength source matching, which have proven difficult in edge-on systems including our Galaxy, NGC~253, and NGC~4945 \citep[][]{Stolte_etal_2014,Emig_etal_2020,Levy_etal_2022}.

\begin{figure*}[t]
\centering
\gridline{
\fig{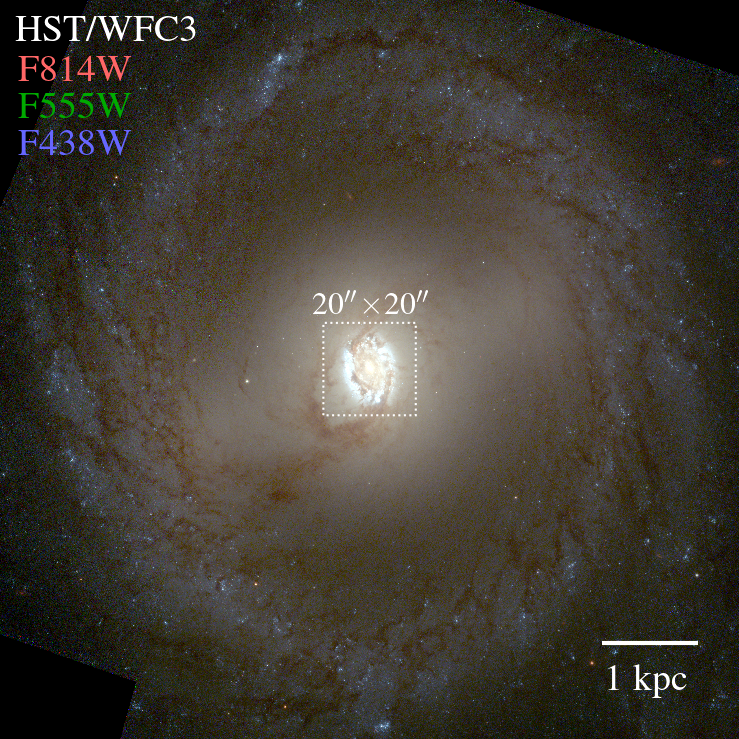}{0.49\textwidth}{}
\hfill
\fig{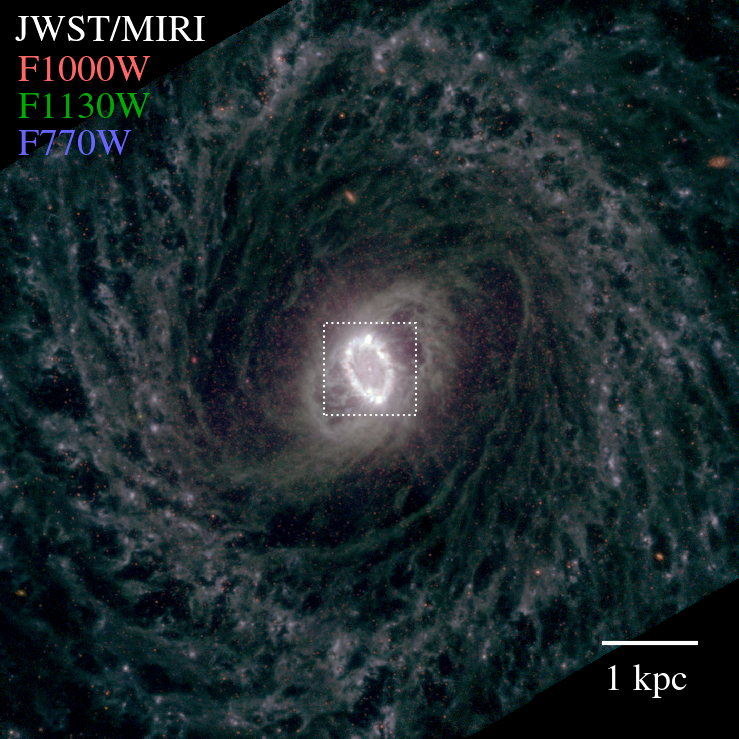}{0.49\textwidth}{}
}
\vspace{-2.0\baselineskip}
\caption{
\textit{Left:} HST/WFC3 composite image of the galaxy NGC~3351 \citep[F814W+F555W+F438W;][]{Lee_etal_2022}.
\textit{Right:} JWST/MIRI composite image for the same field \citep[F1000W+F1130W+F770W;][]{Lee_etal_2023}.
In each panel, a white square marks the central 20''$\times$20'' area, which covers the starburst ring and is roughly the field-of-view of our ALMA observations.}
\label{fig:wide_field}
\end{figure*}

In this paper, we use nearly matched-resolution ALMA, HST, and JWST observations to examine a population of forming YMCs in a nearby Milky Way analog galaxy, NGC~3351 (a.k.a.\ M95, see Table~\ref{tab:M95} for its basic properties).
This galaxy features a prominent, dusty, inner ring structure, which is likely fed by large-scale gas inflows induced by a strong stellar bar \citep[e.g.,][see Figure~\ref{fig:wide_field}]{Regan_etal_2006}.
This ``starburst ring'' hosts a large number of optically-visible massive clusters, and previous IR observations show signs of embedded YMC formation as well \citep[e.g.,][]{Ma_etal_2018,Calzetti_etal_2021,Turner_etal_2021}.
Its favorable viewing angle ($i\approx45^\circ$, Table~\ref{tab:M95}) and clean orbital configurations (Figure~\ref{fig:wide_field}) enable cross-wavelength source matching as well as examinations of systematic trends along the ring.

The structure of this paper is as follows.
Section~\ref{sec:data} describes the ALMA, HST, and JWST datasets used in this paper.
Section~\ref{sec:analyses} details our source identification, characterization, and cross-matching schemes.
Section~\ref{sec:results} presents a set of key physical properties measured for all ALMA-identified YMC candidates.
Section~\ref{sec:discuss} synthesizes the observational results from ALMA, JWST, and HST, constructs an evolutionary timeline for YMC formation, and puts the YMC population in the large-scale context of the starburst ring in NGC~3351.
We summarize all our findings in Section~\ref{sec:conclusion}.

\begin{deluxetable}{rl}
\tablecaption{Basic Properties of the Target Galaxy\label{tab:M95}}
\tablewidth{0pt}
\startdata
\\[-0.5em]
Galaxy name & NGC~3351 (M95) \\
Galaxy type & SB(r)b \\
Center coordinates & ($\rm10^h43^m57.73^s$, $\rm+11^\circ42^\prime13.3^{\prime\prime}$) \\
Distance & $9.96\,(\pm0.01)$~Mpc [1] \\
Systemic velocity & $775\,(\pm5)\rm\;km/s$ [2] \\
Inclination angle & $45\,(\pm6)$~deg [2] \\
Position angle & $193\,(\pm2)$~deg [2] \\
Stellar mass & $2.3\,(\pm0.6){\times}10^{10}\;\uM$ [3] \\
SFR & $1.3\,(\pm0.3)\;\uSFR$ [3] \\
\enddata
\tablecomments{
[1] \citet{Anand_etal_2021}; 
[2] \citet{Lang_etal_2020};
[3] \citet{Leroy_etal_2021a}.}
\vspace{-1.7\baselineskip}
\end{deluxetable}
\vspace{-1\baselineskip}

\section{Data} \label{sec:data}

We use a new set of very high resolution ALMA observations in this work.
Here we lay out the observational design, reduction procedures, and data characteristics for this new ALMA dataset (Section~\ref{sec:data:ALMA}).
We then briefly describe ancillary HST and JWST images and data products used in this work (Section~\ref{sec:data:HST}).

\subsection{ALMA Data} \label{sec:data:ALMA}

We acquired ALMA Band~3 and 7 observations in Cycle~8 (2021.1.00059.S; PI: J.~Sun) to highly resolve the central starburst region of NGC~3351 (Figure~\ref{fig:wide_field} and \ref{fig:alma-zoom}).
These observations capture (sub-)millimeter continuum emission and molecular line emission at $0\farcs1$--$0\farcs2$ resolution (5--10~pc at 9.96~Mpc), which matches the typical size of YMCs \citep[diameter $\sim4$--10~pc;][]{Ryon_etal_2017}.

Our Band~3 observations employ the most extended 12-m array configuration offered in Cycle~8 (C-8) to reach our desired resolution of $0\farcs1$--$0\farcs2$.
These long-baseline data are supplemented by shorter-spacing data acquired in C-5 (same Cycle~8 project) and C-2 configurations \cite[archival data from Cycle~2;][]{Gallagher_etal_2018a} to extend the maximal recoverable scale (MRS) to the size of the entire central starburst region ($\sim20\arcsec$).
The Cycle~8 observations cover a $\sim60\arcsec$ field-of-view (FoV) with a single 12-m pointing and reach on-source integration times of 1.2~h (C-8) and 0.3~h (C-5).
The Band~3 spectral tuning covers the 85--101~GHz continuum and the \HCN10, \HCOp10, and \CS21\ lines at a native velocity resolution of $3.4\;\uV$.

Our Band~7 data combine C-6 and C-3 configurations of the 12-m array and supplementary 7-m array observations (all from the same Cycle~8 project) to achieve a similar resolution and MRS as our Band~3 observations.
The 12-m observations cover a $\sim30\arcsec$ FoV using a 7-pointing mosaic and reach on-source integration times of 0.4~h (C-6) and 0.2~h (C-3); the 7-m observations use a 3-pointing mosaic to cover a similar FoV, with 0.5~h of on-source integration.
The Band~7 spectral tuning covers the 342--357~GHz continuum and the \CO32, \HCOp43, and \CS76\ lines at a native velocity resolution of $0.95\;\uV$.

\begin{deluxetable*}{lccccc}
\tablecaption{ALMA Data Products and Characteristics\label{tab:data}}
\tablewidth{0pt}
\tablehead{
\colhead{Product} &
\colhead{Band} &
\colhead{Arrays} &
\colhead{Beam FWHM} &
\colhead{Channel~Width} & 
\colhead{1$\sigma$ Noise Level (at mosaic center)} \\[-1.5em]
}
\startdata
\\[-1.5em]
\multicolumn{6}{c}{Continuum Images} \\
\hline
93~GHz & 3 & C8+C5+C2 & $0\farcs17$ (8.2~pc) & -- & 6.5\;$\mu$Jy/beam \\
350~GHz & 7 & C6+C3+7m & $0\farcs17$ (8.2~pc) & -- & 83\;$\mu$Jy/beam \\
\hline
\multicolumn{6}{c}{Molecular Line Data Cubes} \\
\hline
\CO32\ & 7 & C6+C3+7m & $0\farcs10$ (4.8~pc) & 0.85\;\uV\ & 1.9\;K \\
\CO32\ & 7 & C6+C3+7m & $0\farcs10$ (4.8~pc) & 5\;\uV\ & 0.9\;K \\
\HCN10\ & 3 & C8+C5+C2 & $0\farcs19$ (9.2~pc) & 10\;\uV\ & 1.1\;K\\
\HCOp10\ & 3 & C8+C5+C2 & $0\farcs19$ (9.2~pc) & 10\;\uV\ & 1.1\;K\\
\CS21\ & 3 & C8+C5+C2 & $0\farcs19$ (9.2~pc) & 10\;\uV\ & 1.0\;K\\
\HCOp43\ & 7 & C6+C3+7m & $0\farcs12$ (5.8~pc) & 10\;\uV\ & 0.4\;K \\
\CS76\ & 7 & C6+C3+7m & $0\farcs12$ (5.8~pc) & 10\;\uV\ & 0.3\;K \\
\enddata
\tablecomments{
The Band~3 C-2 data come from an archival project \citep[2013.1.00634.S;][]{Gallagher_etal_2018a}. All other data are newly acquired in Cycle~8 (2021.1.00059.S; PI: J.~Sun).
}
\vspace{-1.5\baselineskip}
\end{deluxetable*}
\vspace{-2\baselineskip}

\subsubsection{Calibration and Imaging}\label{sec:data:ALMA:imaging}

We calibrate the raw visibility data with observatory-supplied scripts and the appropriate version of CASA pipeline (6.2.1 for Cycle~8).
The calibrated data show no obvious pathologies upon visual inspection.

From the calibrated measurement sets, we extract and image a relevant subset of visibility data for each molecular line and continuum using a modified version of the PHANGS--ALMA imaging pipeline \citep{Leroy_etal_2021b}.
Here, we outline the workflow and highlight a few deviations from the original pipeline.

To prepare for continuum imaging, we extract all line-free channels from the calibrated measurement set and regrid them into a small number of channels per spectral window (SPW).
Here, rather than collapsing each SPW into one channel (default behavior of the PHANGS pipeline), we manually choose a small enough output channel width to prevent bandwidth smearing.
We then combine the continuum-only data from all array configurations (now on a common spectral grid) to make a joint dataset for the continuum in each band.

To prepare for line imaging, we model the continuum emission with a 1st-order polynomial based on all line-free channels across all SPWs of a particular band.
We then subtract this model off from the calibrated measurement set, extract the subset of channels within the velocity range of interest ($\pm300\,\uV$ around the systemic velocity of $778\,\uV$) for each emission line, and regrid the line spectrum to a desired channel width.
Here, we choose a $10\,\uV$ channel width for all high critical density molecular lines (HCN, HCO$^+$, CS) to ensure a reasonable signal-to-noise (S/N) ratio per channel; we keep the native $0.85\,\uV$ channel width for \CO32\ as S/N ratio is not a concern there.
We similarly combine the continuum-subtracted line data from all array configurations (now on a common spectral grid) to make a joint dataset for each emission line.

We image the joint, calibrated visibility dataset for each emission line and continuum following broadly the PHANGS imaging scheme \citep{Leroy_etal_2021b}.
We first run a shallow, multiscale \texttt{tclean} to identify both compact and extended emission across the entire FoV down to $\mathrm{S/N}=4$.
After this step, we run a deeper, singlescale \texttt{tclean} to pick up remaining emission down to $\mathrm{S/N}=1$ for the \CO32\ line and $\mathrm{S/N}=2$ for all other lines and continua.
To avoid divergence, we restrict this second step to within a cleaning mask, which is constructed based on archival, lower-resolution \CO21\ data \citep{Sun_etal_2020a,Leroy_etal_2021a} and only includes $ppv$ locations with significant \CO21\ emission.
In both \texttt{tclean} calls, we weigh the visibility data using the \textit{Briggs} method with a robustness parameter of 1.0, which offers the best trade-off between sensitivity and resolution for our science goal (i.e., discerning individual YMCs).
This \textit{robust} weighting is applied to all continua and lines except for \CO32, which is bright enough to afford a \textit{uniform} weighting that minimizes side-lobes and facilitates high dynamic range imaging.

\subsubsection{Postprocessing and Data Characteristics}\label{sec:data:ALMA:post}

After obtaining the cleaned continuum images and line data cubes, we correct them for the primary beam response pattern and then mildly convolve them so that the beam shapes become round.
We also convert all line data cubes to brightness temperature units (K) according to the corresponding line frequencies.

The output Band~3 and 7 continuum images have circularized beam sizes of $0\farcs17$ and $0\farcs13$, respectively\footnote{The quoted beam sizes are the beam full width at half maximum (FWHM) throughout this paper.}.
To facilitate rigorous comparisons between them, we further convolve the Band~7 continuum image to $0\farcs17$ resolution to match the Band~3 image.
These final continuum images have noise levels of $\sim$6.5--6.8~$\mu$Jy/beam (Band~3) and $\sim$80--100~$\mu$Jy/beam (Band~7) across the central starburst region (also see Table~\ref{tab:data}).

The output data cube for the \CO32\ line reaches a spatial resolution of $0\farcs10$ thanks to the \textit{uniform} weighting. 
The noise level at the native $0.85\;\uV$ channel width is 1.9--2.5~K, which already enables detection of \CO32\ emission across a large number of channels at high significance in most regions.
To further increase the image dynamic range, we also generate another version of the \CO32\ cube by rebinning to a $5\;\uV$ channel width, which lowers the noise down to 0.9--1.2~K and allows us to pick up more diffuse emission (Table~\ref{tab:data}).

The output data cubes for the other high critical density molecular lines [\HCOp43\ and \CS76\ in Band~7; \CS21, \HCOp10, and \HCN10\ in Band~3] have beam sizes of $0\farcs12$--$0\farcs19$.
The noise levels are 0.3--0.5~K (Band~7) and 1.0--1.2~K (Band~3) per 10~$\uV$ channel (Table~\ref{tab:data}).

\begin{figure*}[p]
\gridline{
\fig{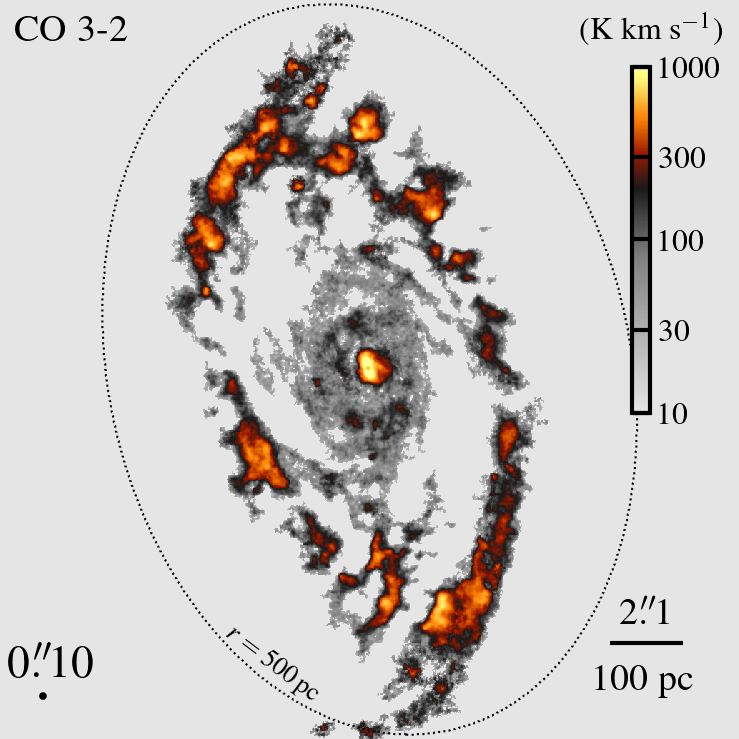}{0.49\textwidth}{}
\hfill
\fig{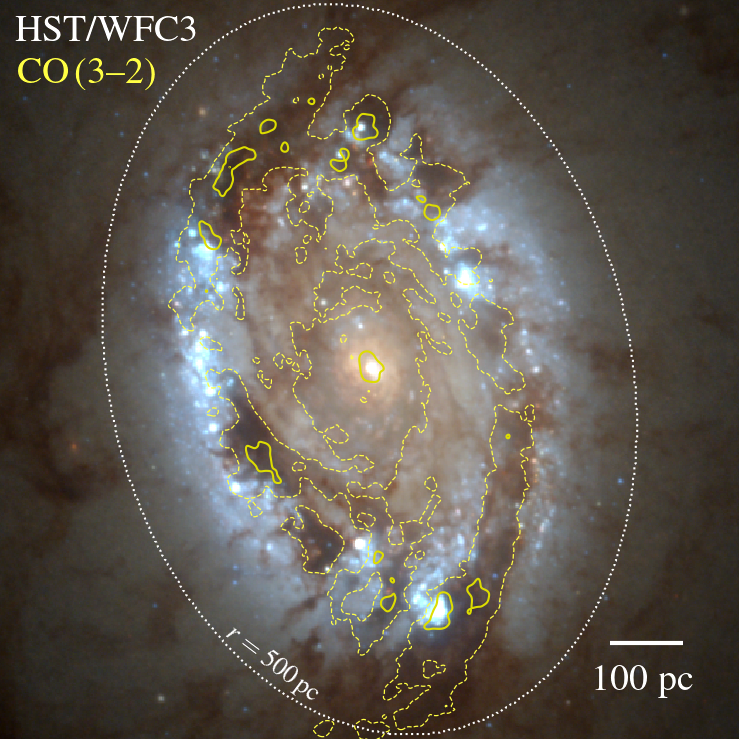}{0.49\textwidth}{}
}
\vspace{-2\baselineskip}
\gridline{
\fig{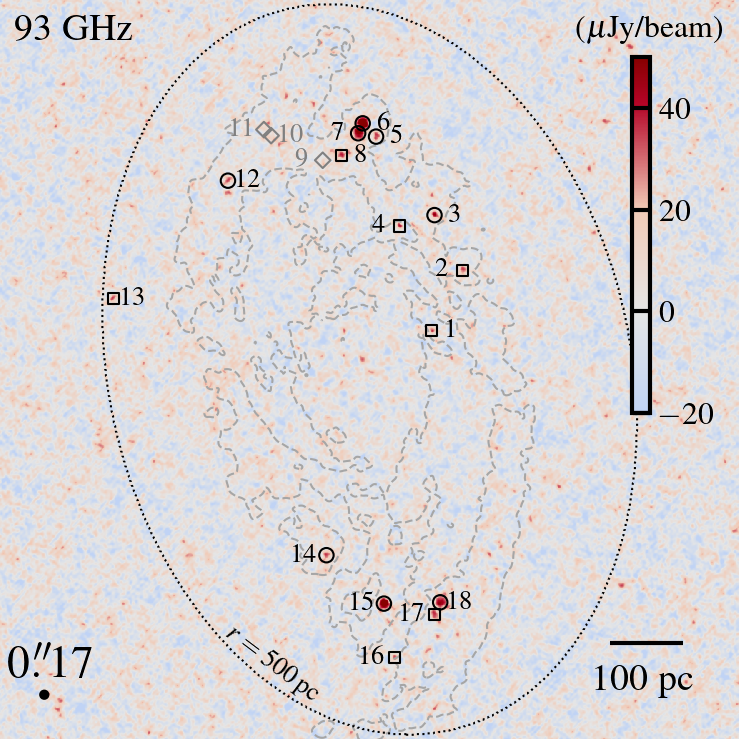}{0.49\textwidth}{}
\hfill
\fig{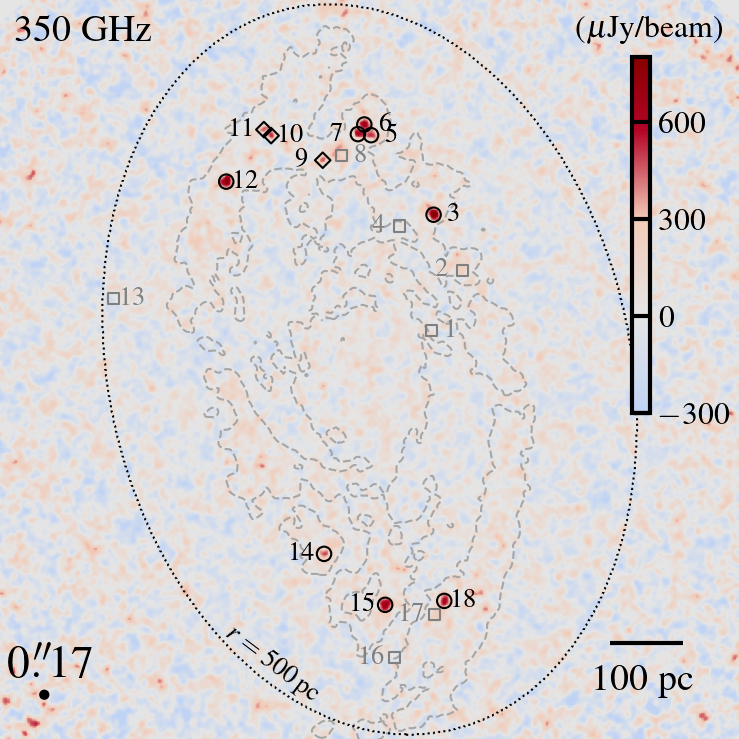}{0.49\textwidth}{}
}
\vspace{-2\baselineskip}
\caption{
\textit{Top left:} ALMA \CO32\ moment-0 map of the central 1$\times$1~kpc ($\sim$20''$\times$20'') of NGC~3351, tracing the distribution of the bulk molecular gas. This moment-0 map is made from the \CO32\ cube with $0\farcs10$ beam size and 5\;\uV\ channel width (see Section~\ref{sec:data:ALMA:post}). The dotted ellipse in the plot corresponds to a galactocentric radius of 500~pc, which encircles the region of interest in this study.
\textit{Top right:} 
HST image for the same area overlaid with \CO32\ integrated intensity contours in yellow (dashed -- 30$\;\uIco$; solid -- 400$\;\uIco$).
Note the clear correspondence between the dust lanes and the CO contours, both of which trace gas distribution.
\textit{Bottom:} ALMA 93~GHz (band~3) and 350~GHz (band~7) continuum images overlaid with only the outer \CO32\ contour (30$\;\uIco$) in light grey. The black circles mark the continuum sources detected in both bands, whereas the square/diamond symbols mark those detected only in band~3/7.
}
\label{fig:alma-zoom}
\end{figure*}

\subsection{HST Data} \label{sec:data:HST}

We make use of multi-band \textit{Hubble Space Telescope} (HST) images of NGC~3351 acquired by the LEGUS \citep{Calzetti_etal_2015} and PHANGS--HST \citep{Lee_etal_2022} surveys.
These are broad-band imaging data obtained with the Wide Field Camera 3 (WFC3), together covering the full galaxy from UV to optical wavelengths (Figure~\ref{fig:wide_field} left panel; also see Figure 1 in \citealt{Turner_etal_2021} for the sky footprint covered by each survey).
The point spread functions (PSFs) of these data are $\lesssim0\farcs1$ in FWHM, slightly better than the beam sizes achieved with our ALMA observations.

To facilitate source cross-matching with ALMA, we take advantage of an updated PHANGS-HST star cluster catalog presented in \citet{Maschmann_etal_2024} and \citet{Thilker_etal_2024}.
This new catalog builds upon the previous PHANGS-HST catalog release \citep[i.e., DR3/CR1\footnote{\url{https://archive.stsci.edu/hlsp/phangs/phangs-cat}} in Feburary~2022; see][]{Lee_etal_2022}.
Aside from improved cluster classification schemes \citep[building on \citealt{Whitmore_etal_2021,Thilker_etal_2022}]{Hannon_etal_2023},
a major improvement is in the spectral energy distribution (SED) fitting, for which a physically motivated ``decision tree'' has been introduced to help address photometric degeneracies between age, extinction, and metallicity \citep[e.g.,][]{Turner_etal_2021,Whitmore_etal_2023a}.
Specifically, this decision tree uses source morphology, broadband color information, and location relative to H$\alpha$-bright complexes (based on ground-based narrowband H$\alpha$ observations; A.~Razza et al., in preparation) to pre-select candidates of young reddened clusters. 
Their SEDs are subsequently fit with a separate set of stellar population models with young ages (${\leq}6$~Myr) and potentially higher extinction (up to $\rm E(B{-}V)=3.0$).
This refined scheme results in more complete identification and better age estimates for young clusters with associated H$\alpha$ emission throughout the PHANGS--HST sample.
We refer interested readers to \citet{Thilker_etal_2024} for more thorough descriptions of the rationale, implementation, and results of this new SED fitting scheme.

In this work, we use this updated cluster catalog to cross-match with sources detected in the ALMA data and compare their estimated physical properties.
Within our region of interest (i.e., the central region of NGC~3351), the new SED fitting scheme gives refined age estimates for ${\sim}65\%$ of clusters.
This fraction is higher than typical due to the dusty and crowded nature of this region, which makes SED fitting with older methods particularly challenging.
Further inspection suggests that the new age estimates are either similar to or more reasonable than the old solutions in ${\sim}80{-}90\%$ of the cases \citep[see Section~6.1 in][]{Thilker_etal_2024}.
The remainder are mostly caused by incorrect association of H$\alpha$ emission to slightly older clusters (due to the much coarser ${\sim}1\farcs2$ PSF of the ground-base H$\alpha$ data used in the decision tree), which affect the accuracy of cluster age estimates primarily between $1{-}10$~Myr.
To mitigate this minor issue, we generally avoid using age estimates for individual clusters and only rely on more robust aggregate statistics, such as the \textit{total number} of clusters younger than $10$~Myr.
We anticipate future works based on higher-resolution HST narrowband H$\alpha$ observations (PIs: F.~Belfiore, R.~Chandar, D.~Thilker) to thoroughly address this issue.

\subsection{JWST Data} \label{sec:data:JWST}

We make use of multi-band JWST NIRCam/MIRI images of NGC~3351 acquired as part of the PHANGS--JWST Cycle 1 Treasury survey \citep[see Figure~\ref{fig:wide_field} right panel]{Lee_etal_2023}.
These broad- and medium-band images are available from the PHANGS DR1.0.1 public image release \citep{Williams_etal_2024}.
This work primarily uses data in the 2.0--11.3~$\micron$ wavelength range, with PSFs ranging from $0\farcs066$ (F200W with NIRCam) to $0\farcs38$ (F1130W with MIRI) in FWHM.

For source cross-matching with ALMA, we also use a catalog of compact 3.35~$\micron$ sources constructed by J.~Rodr\'iguez et al.\ (in preparation) following the methodology of \citet{Rodriguez_etal_2023}.
In short, the sources are identified in the F335M image as local maxima above three times the estimated background level in the neighboring ${<}100$~pc area.
This is done with a peak-finding algorithm \textit{find\_peaks} in \texttt{PHOTUTILS} \citep{Bradley_etal_2022}.
In more crowded regions, the software \texttt{SExtractor} \citep[Source-Extractor;][]{Bertin_Arnouts_1996} is also employed to deblend the sources with a Mexican-hat filter.
This F335M-based source identification scheme allows for more systematic detections of embedded star clusters \citep[see][]{Rodriguez_etal_2023}, which is especially important for cross-matching with ALMA sources in this work.


\section{Analyses} \label{sec:analyses}

From the rich ALMA data set (see Table~\ref{tab:data}), we identify compact (sub-)millimeter continuum sources, characterize their observable properties, and cross-match them with sources in the HST and JWST data.
We detail these analyses in the following subsections.

\subsection{Identifying ALMA Sources}
\label{sec:analyses:identify}

We identify bright, compact sources in the 93~GHz and 350~GHz continuum images as YMC candidates, following \citet{Leroy_etal_2018} and \citet{Emig_etal_2020}.
These sources are selected based on a minimum $\mathrm{S/N}=5$ at peak brightness.
Considering the number of independent measurements (beams) across the region of interest ($r\,{<}\,500$~pc, Figure~\ref{fig:alma-zoom}), this detection threshold corresponds to a false positive rate of $<0.01$, i.e., there is ${<}1\%$ chance for \textit{any} of our selected sources to be spurious.
We choose this rather stringent criterion to ensure sample purity, which is critical for following analysis based on the source number counts (Section~\ref{sec:discuss:timescale}).
This is however at the expense of a higher false negative rate, i.e., missing many potentially real sources with lower peak brightness (as can be seen in Figure~\ref{fig:alma-zoom}).

We identify 15 sources at 93~GHz and 11 at 350~GHz.
Among these, 8 sources at 93~GHz coincide spatially with counterparts at 350~GHz.
These are likely the same objects seen at both frequencies.
Considering these as duplicates, we identify in total 18 ALMA sources, all of which are marked and labeled in Figure~\ref{fig:alma-zoom} (lower panels; also see Table~\ref{tab:YMC_obs}).
Almost all sources appear to be compact and well-isolated, with a single peak and a relatively round shape.
The only exceptions are two closely connected sources in the 93~GHz image (sources \#6 and \#7 in Figure~\ref{fig:alma-zoom}), which we separate by visual inspection.

All the 350~GHz sources and the majority of the 93~GHz sources are located in regions with bright \CO32\ emission (see contours in Figure~\ref{fig:alma-zoom} bottom panels), consistent with forming YMCs surrounded by substantial molecular gas reservoir.
There are however four 93~GHz-only sources (\#1, \#4, \#13, \#16) that are located in regions with no detectable \CO32\ emission.
We will discuss the nature of these sources in Section~\ref{sec:discuss:nature}.

\begin{deluxetable*}{lccccccc}[ht]
\tablecaption{Observed Properties of the ALMA Sources\label{tab:YMC_obs}}
\tablewidth{0pt}
\tablehead{
\colhead{ID} &
\colhead{Coordinate} &
\colhead{$S\sbsc{93}$} &
\colhead{$R\sbsc{hl,\,93}$} &
\colhead{$S\sbsc{350}$} & 
\colhead{$R\sbsc{hl,\,350}$} &
\colhead{$L'\sbsc{HCN\,1-0}$} &
\colhead{$\sigma\sbsc{HCN\,1-0}$} \\
\colhead{} &
\colhead{[$^\circ$]} &
\colhead{[mJy]} &
\colhead{[pc]} &
\colhead{[mJy]} &
\colhead{[pc]} &
\colhead{$[10^3\,\uLco]$} &
\colhead{$[\uV]$} \\
\colhead{(1)} &
\colhead{(2)} &
\colhead{(3)} &
\colhead{(4)} &
\colhead{(5)} &
\colhead{(6)} &
\colhead{(7)} &
\colhead{(8)} \\[-2em]
}
\startdata
\\[-1.5em]
1 & $160.990061,\,+11.703993$ & $0.037\pm0.012$ & $(4.3\pm0.6)$ & $<$$0.33$ & $\cdots$ & $\cdots$ & $\cdots$ \\
2 & $160.989812,\,+11.704463$ & $0.053\pm0.016$ & $2.9\pm1.5$ & $<$$0.34$ & $\cdots$ & $4.7\pm1.5$ & $8.6\pm3.0$ \\
3 & $160.990040,\,+11.704893$ & $0.034\pm0.009$ & $(3.7\pm0.4)$ & $1.23\pm0.22$ & $2.9\pm1.0$ & $7.4\pm1.9$ & $9.5\pm2.9$ \\
4 & $160.990317,\,+11.704807$ & $0.029\pm0.008$ & $(3.3\pm0.3)$ & $<$$0.34$ & $\cdots$ & $\cdots$ & $\cdots$ \\
5 & $160.990521,\,+11.705508$ & $0.058\pm0.016$ & $1.4\pm3.4$ & $1.85\pm0.44$ & $6.6\pm1.4$ & $<$$12.7$ & $\cdots$ \\
6 & $160.990601,\,+11.705601$ & $0.134\pm0.018$ & $3.5\pm0.6$ & $1.87\pm0.37$ & $5.4\pm1.0$ & $23.4\pm4.3$ & $15.8\pm4.2$ \\
7 & $160.990644,\,+11.705525$ & $0.140\pm0.028$ & $6.0\pm1.0$ & $1.29\pm0.30$ & $4.1\pm1.2$ & $12.3\pm3.2$ & $9.2\pm2.8$ \\
8 & $160.990779,\,+11.705358$ & $0.040\pm0.014$ & $(4.6\pm0.7)$ & $<$$0.35$ & $\cdots$ & $6.2\pm1.9$ & $7.9\pm2.8$ \\
9 & $160.990925,\,+11.705318$ & $<$$0.033$ & $\cdots$ & $0.98\pm0.29$ & $4.1\pm1.2$ & $9.9\pm2.8$ & $11.6\pm4.3$ \\
10 & $160.991335,\,+11.705510$ & $<$$0.033$ & $\cdots$ & $0.81\pm0.23$ & $2.6\pm1.1$ & $<$$5.5$ & $\cdots$ \\
11 & $160.991393,\,+11.705557$ & $<$$0.033$ & $\cdots$ & $1.07\pm0.32$ & $4.0\pm1.6$ & $<$$6.5$ & $\cdots$ \\
12 & $160.991685,\,+11.705156$ & $0.035\pm0.013$ & $(4.5\pm0.7)$ & $2.72\pm0.40$ & $6.2\pm0.8$ & $15.0\pm3.6$ & $10.6\pm3.1$ \\
13 & $160.992590,\,+11.704243$ & $0.041\pm0.014$ & $(4.5\pm0.7)$ & $<$$0.37$ & $\cdots$ & $\cdots$ & $\cdots$ \\
14 & $160.990905,\,+11.702250$ & $0.056\pm0.017$ & $2.9\pm1.3$ & $0.62\pm0.24$ & $3.3\pm1.4$ & $<$$6.5$ & $\cdots$ \\
15 & $160.990434,\,+11.701863$ & $0.152\pm0.020$ & $4.0\pm0.6$ & $3.17\pm0.48$ & $7.6\pm0.9$ & $20.7\pm5.5$ & $8.6\pm2.6$ \\
16 & $160.990357,\,+11.701448$ & $0.026\pm0.008$ & $(3.4\pm0.4)$ & $<$$0.37$ & $\cdots$ & $\cdots$ & $\cdots$ \\
17 & $160.990040,\,+11.701785$ & $0.083\pm0.022$ & $4.2\pm1.7$ & $<$$0.36$ & $\cdots$ & $10.5\pm2.6$ & $12.0\pm3.9$ \\
18 & $160.989975,\,+11.701884$ & $0.153\pm0.028$ & $5.9\pm0.9$ & $1.44\pm0.35$ & $5.3\pm1.3$ & $<$$8.2$ & $\cdots$ \\
\enddata
\tablecomments{
(1) source ID as appeared in Figure~\ref{fig:alma-zoom};
(2) source RA and Dec coordinates;
(3) continuum flux density at 93~GHz;
(4) half-light radius at 93~GHz, defined as the geometric mean of the beam-deconvolved semi-major and semi-minor axes (for sources with unsuccessful beam deconvolution, we report the original values in parentheses);
(5) continuum flux density at 350~GHz;
(6) half-light radius at 350~GHz (note that all detected sources have successful beam deconvolution);
(7) \HCN10\ line integrated luminosity (sources without CO~3--2 emission appear as ``$\cdots$'' due to the lack of prior information for determining velocity range; \S\ref{sec:analyses:characterize});
(8) \HCN10\ line velocity dispersion (i.e., effective width; \S\ref{sec:analyses:characterize}).
}
\vspace{-1\baselineskip}
\end{deluxetable*}
\vspace{-2\baselineskip}

\subsection{Characterizing ALMA Sources}
\label{sec:analyses:characterize}

We measure fluxes and sizes of the 18 sources in the 93~GHz and 350~GHz continua using the CASA task \texttt{imfit}.
This task fits a 2D Gaussian function to the flux distribution within a user-defined region.
First, we define a circular aperture around each identified source with a diameter of $0\farcs34$, which is twice the beam FWHM\footnote{For clustered sources (\#5, \#6, \#7 and \#10, \#11), we manually draw a bigger aperture that encircles all the clustered sources and ask \texttt{imfit} to perform a multi-component Gaussian fitting.}.
We then perform the \texttt{imfit} task on the continuum image within the circular aperture for each source.
Note that we set the \texttt{imfit} parameter \texttt{dooff=True} in order to fit for and subtract a smooth background that may be present at the location of some sources.
We also ask \texttt{imfit} to try deconvolving the beam size from the best-fit 2D Gaussian profile so that we can recover the intrinsic size of each source.
We have visually inspected all \texttt{imfit} results to ensure reliability.

From the \texttt{imfit} results, we record the measured fluxes and radii (original or deconvolved) for all sources and list them in Table~\ref{tab:YMC_obs}.
The source fluxes range over $S_{93}=0.03{-}0.15$~mJy at 93~GHz and $S_{350}=0.6{-}1.9$~mJy at 350~GHz.
The beam-deconvolved source half-light radii (calculated as the geometric mean of the semi-major and semi-minor axes) ranges over $R\sbsc{hl,\,93}=1.4{-}6.0$~pc and $R\sbsc{hl,\,350}=2.6{-}7.6$~pc.
We note that the deconvolution is unsuccessful for several 93~GHz sources because their observed major/minor axis lengths are smaller than the beam size, whereas none of the 350~GHz sources has similar issues.
For the unsuccessful cases, we report the original 2D Gaussian sizes in Table~\ref{tab:YMC_obs} and treat these numbers as upper limits on the intrinsic source radii.

In addition to the continuum measurements, we also extract for each source a set of molecular line spectra (see Table~\ref{tab:data} for a list of included lines).
We define elliptical apertures for spectra extraction based on the best-fit 2D Gaussian for the continuum sources.
For sources with both 93~GHz and 350~GHz detections, we use the apertures determined from the 350~GHz detections, as we expect molecular line emission to be more spatially associated with thermal dust emission than free-free emission.

We also perform background subtraction on the extracted line spectra.
To do this, we define a ``background annulus'' around the source, calculate the mean emission spectrum within that annulus, and subtract this mean spectrum from the extracted spectrum at the location of the source.
For isolated sources, the background annuli are set to have inner/outer radii of $2\times$/$3\times$ the aperture radii for source extraction.
For clustered sources, we manually draw an annulus safely enclosing all the sources and set the outer radius to be $2\times$ the inner radius.
For source \#2 that is in between two \CO32\ blobs (Figure~\ref{fig:alma-zoom}), we omit background subtraction since there is no obvious way to reliably determine the local background level in this case, and that level can be zero for some choices of the background aperture.

After extracting the emission line spectra and subtracting the local background, we measure the line integrated luminosities and line widths for all sources and all lines.
We use the brightest line in our sample, \CO32, as a prior to determine the velocity range of each source and then measure line integrated flux and effective width \citep[with the latter corrected for broadening due to finite channel width; see][]{Heyer_etal_2001,Sun_etal_2018} for all high critical density lines.
This work mainly uses measurements for the \HCN10\ line, which we report in Table~\ref{tab:YMC_obs}.
The spectra of all molecular lines are presented in Appendix~\ref{apdx:spectra}.

\subsection{Cross-matching ALMA--JWST--HST Sources}
\label{sec:analyses:crossmatch}

\begin{figure}[t]
\centering
\gridline{
\fig{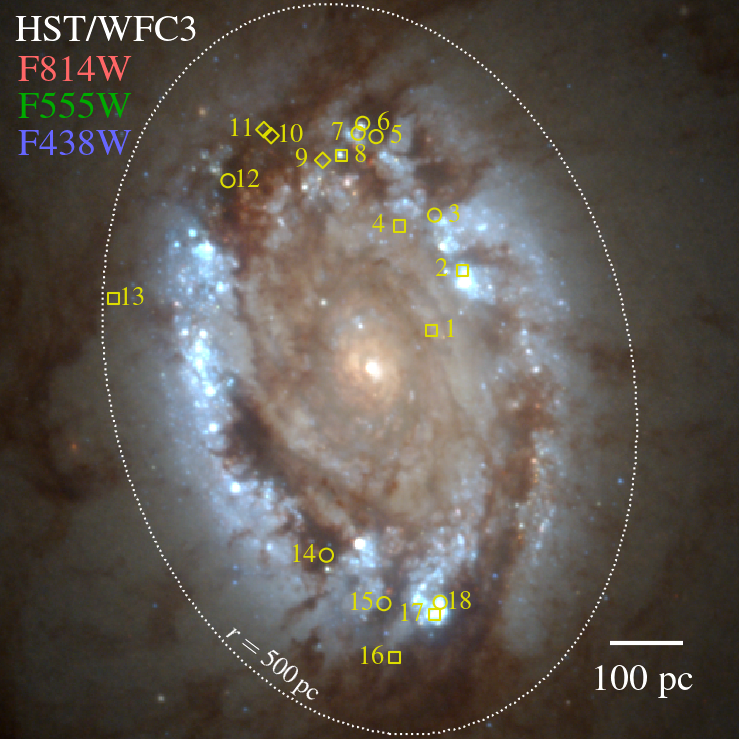}{0.49\textwidth}{}
}
\vspace{-2\baselineskip}
\caption{
HST composite image of the central 1$\times$1~kpc of NGC~3351, with locations of the 18 ALMA sources labeled by yellow symbols.
Sources \#2, \#6, \#7, \#8, \#17, and \#18 have cross-matched HST clusters, with \#6 and \#7 matched to the same cluster (also see Table~\ref{tab:crossmatch}).
}
\label{fig:hst-zoom}
\end{figure}

\begin{figure*}[htb]
\centering
\gridline{
\fig{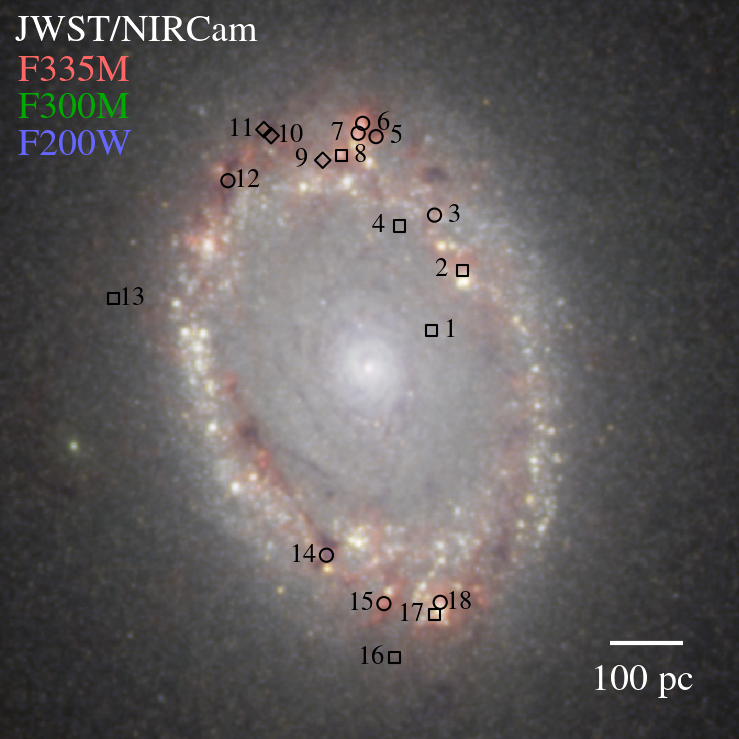}{0.49\textwidth}{}
\hfill
\fig{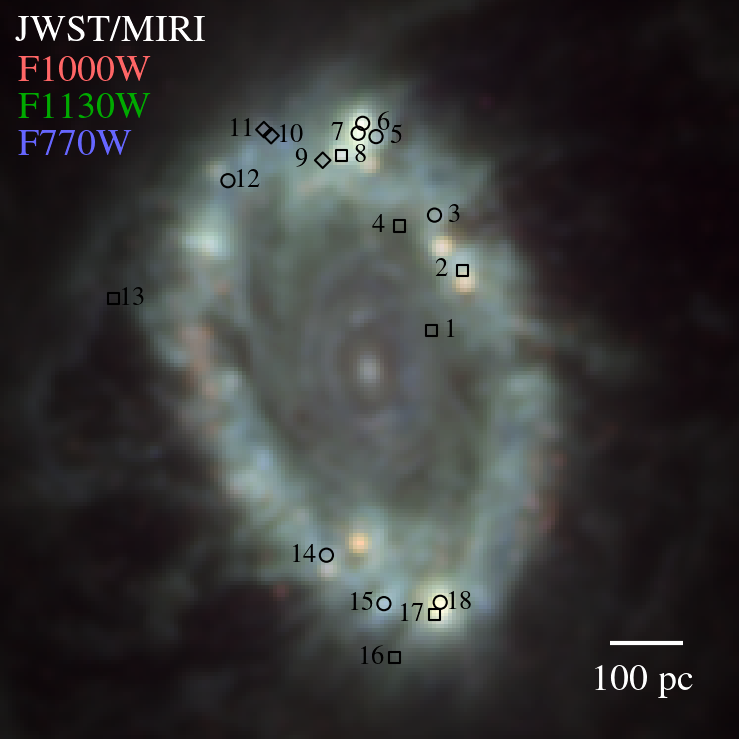}{0.49\textwidth}{}
}
\vspace{-2\baselineskip}
\gridline{
\fig{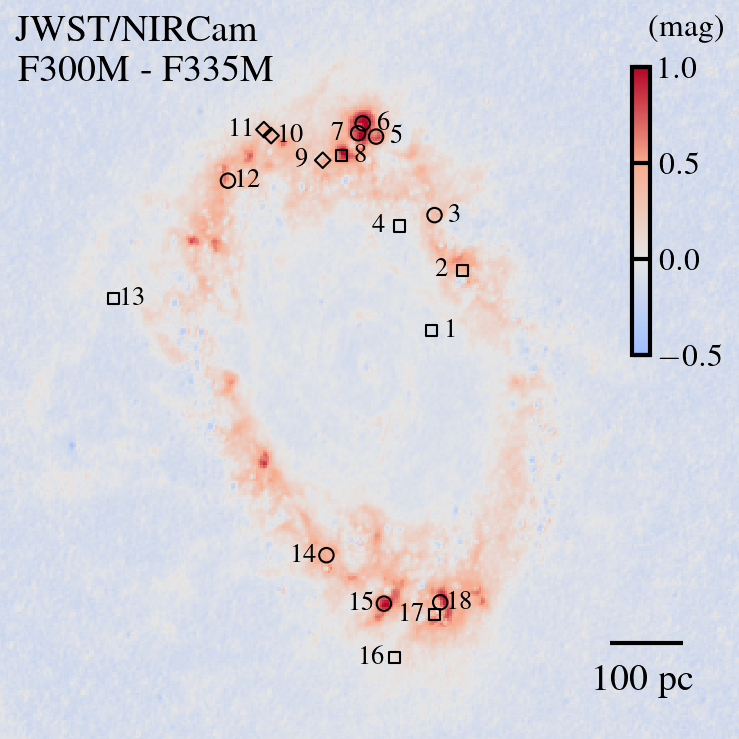}{0.49\textwidth}{}
\fig{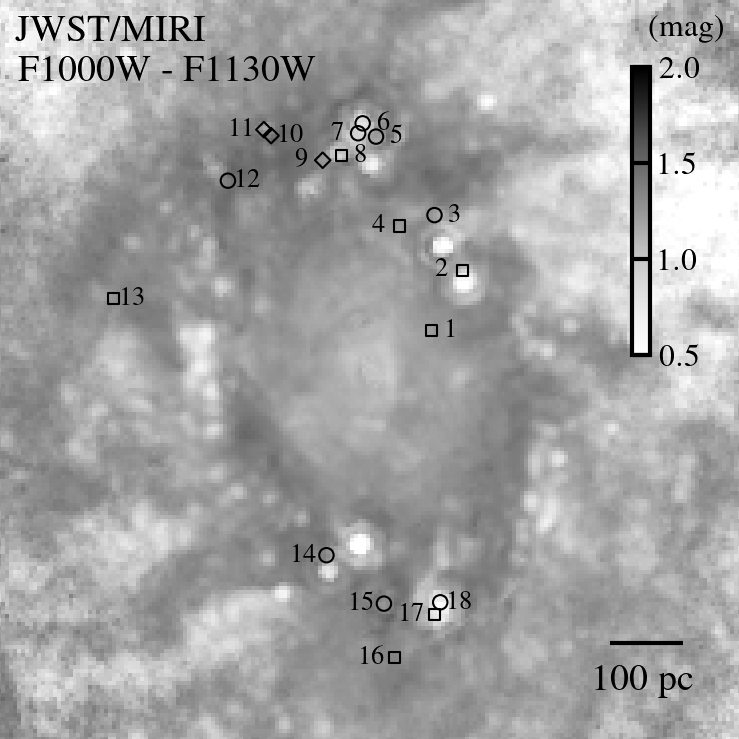}{0.49\textwidth}{}
}
\vspace{-2\baselineskip}
\caption{
JWST NIRCam and MIRI images of the central 1$\times$1~kpc, with locations of ALMA sources marked by black symbols.
\textit{Top left:} NIRCam F335M+F300M+F200W composite image showing the near-IR SED.
\textit{Bottom left:} NIRCam F300M--F335M color image highlighting the 3.35$\micron$ PAH emission.
These two panels are useful for probing young, embedded clusters \citep{Rodriguez_etal_2023}.
\textit{Top right:} MIRI F1000W+F1130W+F770W composite image showing the mid-IR PAH and hot dust emission.
\textit{Bottom right:} MIRI F1000W--F1130W color image highlighting the PAH to continuum contrast.
These two panels allow for identifying compact hot dust sources associated with \HII\ regions and/or powered by clusters \citep{Hassani_etal_2023}.
We find 8 ALMA sources with potential JWST counterparts based on either cross-matching to the 3.35~$\micron$ source catalog and/or visual inspection of the images (see Section~\ref{sec:analyses:crossmatch} and Table~\ref{tab:crossmatch}).
}
\label{fig:jwst-zoom}
\end{figure*}

\begin{deluxetable}{lcccc}[t]
\tablecaption{Source Cross-matching Results \label{tab:crossmatch}}
\tablewidth{0pt}
\tablehead{
\colhead{ID} &
\colhead{ALMA} &
\colhead{ALMA} &
\colhead{JWST} & 
\colhead{HST} \\[-0.8em]
\colhead{} &
\colhead{350~GHz} &
\colhead{93~GHz} &
\colhead{3.35~$\micron$} & 
\colhead{clusters} \\[-1.5em]
}
\startdata
\\[-1.5em]
1 & & $\circ$ & & \\
2 & & $\circ$ & $\circ$ & $\circ$ \\
3 & $\circ$ & $\circ$ & & \\
4 & & $\circ$ & & \\
5 & $\circ$ & $\circ$ & & \\
6 & $\circ$ & $\circ$ & $\circ$ & \multirow{2}{*}{\rotatebox[origin=c]{90}{$\infty$}} \\
7 & $\circ$ & $\circ$ & $\circ$ & \\
8 & & $\circ$ & $\circ$ & $\circ$ \\
9 & $\circ$ & & & \\
10 & $\circ$ & & & \\
11 & $\circ$ & & & \\
12 & $\circ$ & $\circ$ & & \\
13 & & $\circ$ & & \\
14 & $\circ$ & $\circ$ & $\circ$ & \\
15 & $\circ$ & $\circ$ & $\circ$ & \\
16 & & $\circ$ & & \\
17 & & $\circ$ & $\circ$ & $\circ$ \\
18 & $\circ$ & $\circ$ & $\circ$ & $\circ$ \\
\enddata
\tablecomments{
Meaning of various markers used above:\\
$\circ$ -- unique, unambiguous cross-match;\\
\rotatebox[origin=c]{90}{$\infty$} -- a close ALMA source pair (\#6 and \#7) cross-matched to the same HST cluster.
}
\vspace{-1.5\baselineskip}
\end{deluxetable}

\vspace{-1\baselineskip}

For the 18 sources detected in the (sub-)millimeter ALMA observations, we further check if they have counterparts in the UV--optical HST images and the near- to mid-IR JWST images.
The presence or absence of each source at different wavelengths can inform us the presence or absence of various components in a forming cluster (e.g., gas versus stars) as well as the level of dust extinction towards that cluster.
Such qualitative information is especially important for inferring the time evolution of YMCs, which we will discuss in depth below in Section~\ref{sec:discuss:nature}.

For ALMA--HST source cross-matching, we take the star cluster catalog generated from the multi-band HST images \citep[Section~\ref{sec:data:HST}; also see][and references therein]{Lee_etal_2022} and focus on class 1 and 2 objects (i.e., compact clusters).
As the HST images have higher resolution (sharper PSF) than the ALMA continuum images, we consider an HST cluster to be cospatial with an ALMA source if the distance between their central coordinates is smaller than the major axis half width at half maximum (HWHM) of the ALMA source.
This cross-matching scheme yields six ALMA sources with HST cluster counterparts (Figure~\ref{fig:hst-zoom} and Table~\ref{tab:crossmatch}). 
These HST clusters have SED fitting-based mass estimates ranging over $5{\times}10^4{-}4{\times}10^5\,\uM$ and ages ranging over 1--4~Myr, though the ages for some clusters may be underestimated for the reasons described in Section~\ref{sec:data:HST}.

For ALMA--JWST source cross-matching, we first rely on the 3.35~$\micron$ source catalog \citep[see Section~\ref{sec:data:JWST} and][]{Rodriguez_etal_2023}.
Similar to the ALMA--HST source matching criterion, the center-to-center distance between a JWST 3.35~$\micron$ source and an ALMA source needs to be within the major axis HWHM of the ALMA source to be considered a match.
For ALMA sources with multiple matched 3.35~$\micron$ sources, we use the one with a smaller separation.

One caveat with using only the 3.35~$\micron$ source catalog for cross-matching is that it misses a small number of sources that are only visible in other IR bands, partly due to different PSF sizes and sensitivity.
To address this issue, we also visually inspect the multi-band JWST images (Figure~\ref{fig:jwst-zoom}) near the location of each ALMA source to determine (1) if an IR counterpart is clearly visible in other bands but not included in the 3.35~$\micron$ source catalog, and (2) if a cross-matched 3.35~$\micron$ source is consistently present across other IR bands.
We find the F200W data particularly helpful in this regard, thanks to its smaller PSF (FWHM 0\farcs066 versus 0\farcs11) and longer integration time (1200s versus 390s).
Combining the catalog matching and visual inspection procedures, we find in total eight unique cross-matched sources (Table~\ref{tab:crossmatch}).


\section{Physical Properties of the Young Massive Cluster Candidates} \label{sec:results}

From the measured YMC candidate sizes and flux densities in the ALMA (sub-)millimeter continua, their molecular line fluxes and line widths, as well as ancillary information available for their HST and JWST counterparts (if any), we estimate their gas masses (Section~\ref{sec:results:Mgas}), stellar masses (Section~\ref{sec:results:Mstar}), sizes (Section~\ref{sec:results:size}), and other key properties (Sections~\ref{sec:results:Mtot_fgas}--\ref{sec:results:Sigmagas}).
We detail the derivations below and summarize key quantitative results in Table~\ref{tab:YMC_phys}.

\subsection{Gas Mass} \label{sec:results:Mgas}

\begin{figure}[t]
\centering
\gridline{\fig{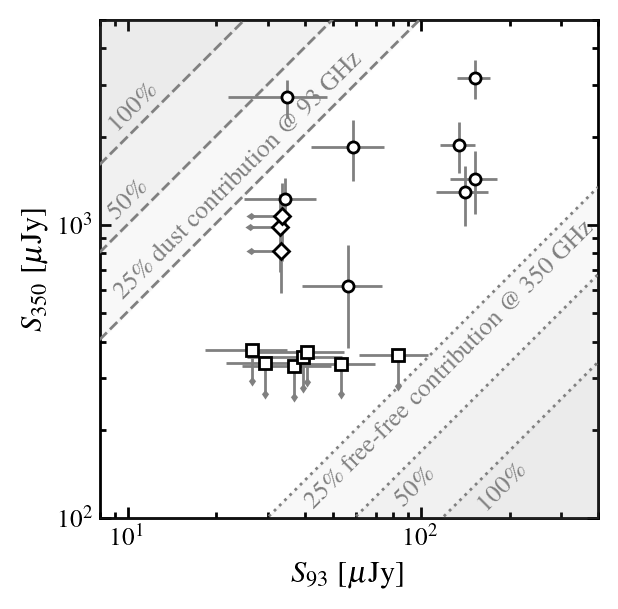}{0.45\textwidth}{}}
\vspace{-2.5\baselineskip}
\caption{Flux densities at 93~GHz versus that at 350~GHz for all identified continuum sources (the symbols are the same as in Figure~\ref{fig:alma-zoom}).
The dashed and dotted lines mark two series of expected relations assuming spectral indices of $\alpha=-0.15$ for free-free and $\alpha=4.0$ for dust.
Almost all sources lie in the unshaded middle region, which suggests that their 93~GHz emission is dominated by free-free and their 350~GHz emission dominated by thermal dust contribution.}
\label{fig:check_flux}
\end{figure}

We use the 350~GHz continuum emission as a dust tracer, from which we infer the masses and sizes of the YMC gas reservoirs.
Generally speaking, the 350~GHz continuum includes not only thermal dust emission, but also free-free and potentially synchrotron emission.
To estimate the fractional contribution of these components, we assume the 93~GHz and 350~GHz flux densities for each object are a mixture of thermal dust and free-free emission, with intrinsic spectral indices of $\alpha=4.0$ (dust) and $\alpha=-0.15$ (free-free), consistent with what previous studies adopted for similar systems \citep[and references therein]{Emig_etal_2020}.
With this simple, two-component decomposition, we derive the fractional contribution of dust and free-free emission at each frequency and find that thermal dust contribution indeed dominates the 350~GHz continuum for all identified objects (see Figure~\ref{fig:check_flux}).
This conclusion agrees with previous YMC studies in other galaxies \citep[e.g.,][]{Leroy_etal_2018,Emig_etal_2020}, for which the free-free contribution to the 350~GHz continuum was also found to be $<10$\% in most cases.

The above analysis ignores the contribution of synchrotron emission, which may be non-negligible at 93~GHz \citep[as found for some YMCs in other systems; see][]{Emig_etal_2020,Mills_etal_2021}.
Qualitatively, we expect the inclusion of a synchrotron component to lower the estimated fraction of free-free at 93~GHz and raise the fraction of thermal dust at 350~GHz.
On the one hand, this makes it even more reasonable to assume the 350~GHz continuum is dust-dominated for all our sources.
On the other hand, it becomes an issue when interpreting the 93~GHz continuum, as discussed below in Section~\ref{sec:results:Mstar}.

After verifying that the 350~GHz continuum is dominated by thermal dust emission, we convert the 350~GHz flux density into a gas mass in two steps.
First, we calculate the dust optical depth at 350~GHz via
\begin{align}
\tau_{350} \approx \frac{S_{350}}{\Omega_{350}\,B_\nu(T\sbsc{dust})}~.
\label{eq:tau350}
\end{align}
\noindent Here $\Omega_{350}$ is the solid angle extended by the 350~GHz source on the sky, and $B_\nu(T\sbsc{dust})$ the blackbody function for a given dust temperature $T\sbsc{dust}$.
We use a dust temperature of $T\sbsc{dust}=130$~K following \citet{Leroy_etal_2018}, but there can be a $\sim$50\% uncertainty on this value.
We note that Equation~\ref{eq:tau350} is a good approximation only in the optically thin limit. Nonetheless, our estimated optical depths of $\tau_{350}\approx0.001\text{--}0.005$ for all sources make this approximation appropriate.

We then estimate the gas mass associated with each source via
\begin{align}
M\sbsc{gas} = \Sigma\sbsc{gas}\,A\sbsc{350} = \frac{\tau_{350}}{\kappa_{350}\,\mathrm{D/G}}\,\Omega_{350}\,d^2~.
\label{eq:Mgas}
\end{align}
\noindent Here $\kappa_{350}=1.9\;\mathrm{cm^2/g}$ is the adopted dust opacity at 350~GHz \citep[appropriate for dust mixed with gas at densities of ${\sim}10^{5{-}6}\,\mathrm{cm^{-3}}$]{Ossenkopf_Henning_1994}, and $\mathrm{D/G}=1/100$ the adopted dust-to-gas ratio \citep{Draine_etal_2007}.
Both values are chosen for consistency with previous YMC studies \citep[e.g.,][]{Leroy_etal_2018,Emig_etal_2020}, but they can introduce systematic uncertainties on the order of $\sim$0.3~dex.
$A\sbsc{350}=\Omega_{350}\,d^2$ is the physical area of the 350~GHz source, which is needed to convert gas surface density into a total gas mass.
The $\Omega_{350}$ term effectively cancels out when combining Equations~\ref{eq:tau350} and \ref{eq:Mgas}.

We find gas masses of $3\times10^4{-}2\times10^5\;\uM$ for the 11 YMCs detected at 350~GHz.
For the non-detections, we put 5$\sigma$ upper limits of $\approx2\times10^4\;\uM$ (see Table~\ref{tab:YMC_phys}).
These gas masses are consistent with expectations for YMC progenitors or the gas reservoir associated with forming YMCs \citep[e.g.,][]{PortegiesZwart_etal_2010}.
Considering uncertainties on the adopted parameters in Equations~\ref{eq:tau350} and \ref{eq:Mgas}, we expect an overall systematic uncertainty of $\sim$0.5~dex on these gas mass estimates.

\begin{figure}[tb]
\centering
\gridline{\fig{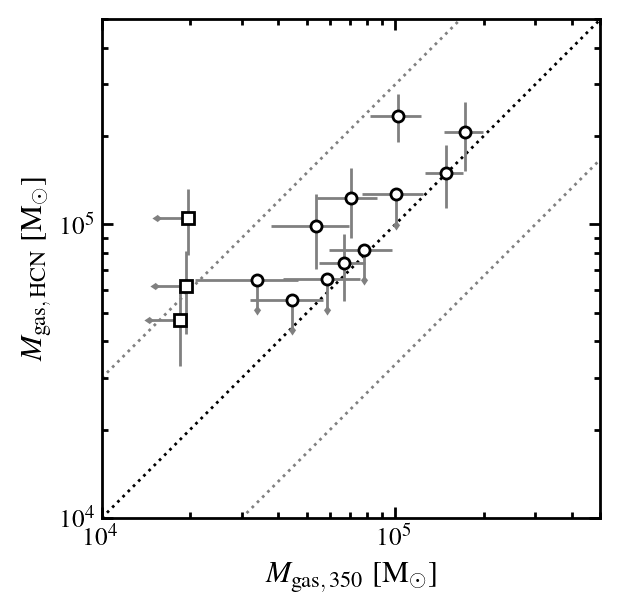}{0.48\textwidth}{}}
\vspace{-2.5\baselineskip}
\caption{
Gas mass estimates from 350~GHz continuum versus those from \HCN10\ line (symbols are the same as in Figure~\ref{fig:check_flux}), with the black and gray dotted lines marking the identity relation and a factor of three offset to either side.
In addition to statistical uncertainties shown by the error bars, these mass estimates are also subject to ${\sim}0.5$~dex systematic uncertainties stemming from our assumptions of dust temperature, emissivity, and abundance as well as HCN conversion factor (\S\ref{sec:results:Mgas}).
The two gas mass estimates broadly agree within the range allowed by those uncertainties.
}
\vspace{-0.5\baselineskip}
\label{fig:check_Mgas}
\end{figure}

As an independent sanity check, we also derive alternative gas masses from the \HCN10\ line luminosities (Section~\ref{sec:analyses:characterize} and Table~\ref{tab:YMC_obs}) with a commonly-adopted HCN-to-H$_2$ conversion factor of $\alpha\sbsc{HCN}\approx10\,\ualphaCO$ \citep{Gao_Solomon_2004}, although lower values have been advocated for systems with more extreme conditions \citep[c.f.\ \citealt{Barnes_etal_2020a}]{Garcia-Burillo_etal_2012,Shimajiri_etal_2017} that in part resemble the local conditions in YMCs.
Our HCN-based gas masses show clear correlation and broad agreement with the 350~GHz-based estimates (Figure~\ref{fig:check_Mgas}). 
Although the former tend to yield slightly higher values, the observed discrepancy is attributable to the large systematic uncertainties on the adopted $\alpha\sbsc{HCN}$ and several coefficients in Equations~\ref{eq:tau350} and \ref{eq:Mgas}.

\subsection{Stellar Mass} \label{sec:results:Mstar}

We use the 93~GHz continuum to constrain the free-free emission from ionized gas in \HII\ regions, from which we infer the ionizing photon production rate and subsequently the mass of the stellar population associated with the YMC candidates.
As mentioned in Section~\ref{sec:results:Mgas}, the 93~GHz continuum may also include contribution from thermal dust and synchrotron emission.
Our simple two-component modelling (dust+free-free; Figure~\ref{fig:check_flux}) suggests that the dust contribution at 93~GHz is less than $20\%$ for all but one source (\#12).
However, the issue of ignoring synchrotron emission in the modelling becomes more concerning at 93~GHz, as synchrotron may have a non-trivial contribution to the observed 93~GHz flux density (e.g., \citealt{Emig_etal_2020} reported a median synchrotron fraction of 0.36 for YMCs in NGC~4945, \citealt{Mills_etal_2021} found similar values for those in NGC~253).

It is possible to better constrain the synchrotron fraction at 93~GHz given sensitive, lower frequency observations at similar angular resolution.
For NGC~3351, previous observations at 1.4~GHz with the Multi-Element Radio Linked Interferometer Network (MERLIN) provide the best lever arm for this purpose \citep{Hagele_etal_2010}.
With no point source detected in the central region at $0\farcs29\times0\farcs17$ resolution, they put a $6\sigma$ upper limit of $0.30\,\uS$ for any compact source at 1.4~GHz.
Adopting a characteristic spectral index of $-0.75$ for synchrotron emission, this translates to an upper limit of $\sim$0.013~$\uS$ at 93~GHz.
In this case, the synchrotron fraction is constrained to be $\lesssim10\%$ for the brightest sources at 93~GHz ($S_{93}\approx0.15\,\uS$) and $\lesssim40\%$ for the fainter ones ($S_{93}\approx0.03\,\uS$, Table~\ref{tab:YMC_obs}).

Considering the lack of more sensitive low-frequency data to further constrain the synchrotron spectral index and fraction, we choose to use our observed 93~GHz flux densities directly as free-free flux densities, which may introduce a bias of at most $40\%$ per our estimates.
We convert the 93~GHz free-free flux densities into ionizing photon production rates, following \citet{Emig_etal_2020}:
\begin{equation}
Q_0 \approx 
1.6\times10^{52}\,\mathrm{s}^{-1}
\left(\frac{S_{93}}{\uS}\right)
\left(\frac{d}{10\,\mathrm{Mpc}}\right)^2
\left(\frac{T\sbsc{e}}{6000\,\mathrm{K}}\right)^{-0.51}~.
\label{eq:Q0}
\end{equation}
\noindent Here we use a characteristic electron temperature of $T\sbsc{e}=6000\,\mathrm{K}$ that is typical of galaxy centers \citep[e.g.,][]{Bendo_etal_2016,Emig_etal_2020,Henshaw_etal_2023}.
Equation~\ref{eq:Q0} also implicitly assumes that the escape fraction of ionizing photons from the YMCs are negligible, which should be a reasonable assumption for the young, deeply embedded clusters studied in this work.

We further infer a stellar mass from the ionizing photon production rate via
\begin{equation}
M\sbsc{\star} \approx \frac{Q_0}{2.2\times10^{46}\,\mathrm{s}^{-1}}\,\uM~.
\label{eq:Mstar}
\end{equation}
\noindent This conversion is based on a Starburst99 simulation \citep{Leitherer_etal_1999} of a 3~Myr old stellar population with a \citet{Kroupa_2001} initial mass function and solar metallicity \citep{Williams_etal_2022}.
If we had assumed a zero-age stellar population \citep[see][]{Leroy_etal_2018}, the estimated masses would be 2$\times$ smaller; an age of 5~Myr would instead yield 5$\times$ larger values \citep[see][]{Emig_etal_2020}.
Our adopted 3~Myr is consistent with the 1--4~Myr age range given by UV--optical photometric SED fitting for the few sources with HST counterparts (Section~\ref{sec:analyses:crossmatch}, modulo caveats described in Section~\ref{sec:data:HST}).
Such young ages seem reasonable since most 93~GHz sources are deeply embedded and have substantial gas reservoirs around them (Section~\ref{sec:results:Mgas}), which suggests that supernova explosions have not gone off (i.e., age $\lesssim4$~Myr).
The adopted 3~Myr is also consistent with our inferred YMC evolutionary timeline below in Section~\ref{sec:discuss:timescale}.

With Equations~\ref{eq:Q0} and \ref{eq:Mstar}, we find stellar masses of $2\times10^4{-}1\times10^5\;\uM$ for the 15 YMCs with detectable 93~GHz emission.
For non-detections, the 5$\sigma$ upper limit on the 93~GHz flux translates to $\approx2.5\times10^4\;\uM$ under the same assumptions (see Table~\ref{tab:YMC_phys}).
These stellar mass estimates are consistent with typical definition of YMCs \citep[or super star clusters, SSCs;][]{PortegiesZwart_etal_2010}.
The main source of systematic uncertainty is the adopted stellar age for Equation~\ref{eq:Mstar}, which can introduce 0.3--0.6~dex of variations as discussed above.

For the few 93~GHz sources with HST counterparts, we find that the 93~GHz-based stellar mass estimates are systematically lower than the HST SED-based mass estimates (median offset ${\sim}0.5$~dex, Table~\ref{tab:YMC_phys}).
On the one hand, these sources are less embedded (visible in HST data), so it is possible that (1) they tend to be slightly older than 3~Myr on average, and (2) a non-negligible fraction of the ionizing photons do escape from the clusters. 
Both of these would lead to underestimated stellar masses via Equation~\ref{eq:Q0}.
On the other hand, the UV--optical SED-based mass estimates also have non-negligible uncertainties as discussed in Section~\ref{sec:data:HST}.
We aim to address this issue in a follow-up study by modelling the full UV--optical--IR SED for all clusters identified in the joint HST+JWST dataset, potentially including the ALMA measurements whenever available.

\subsection{Size}
\label{sec:results:size}

\begin{figure}[b]
\gridline{
\fig{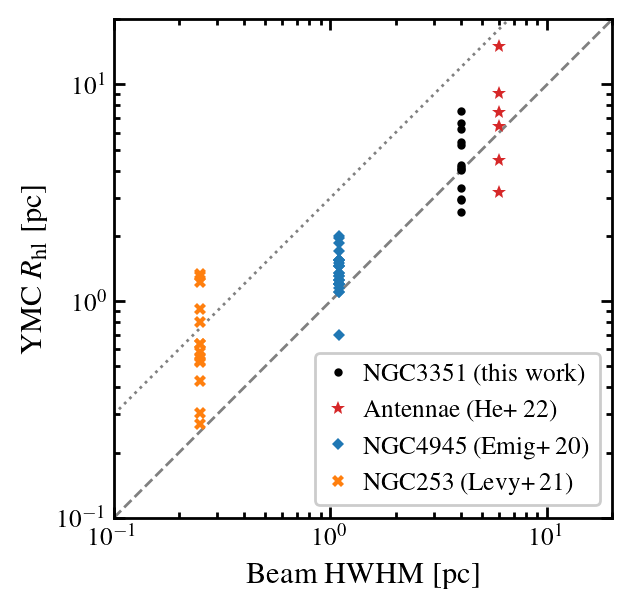}{0.48\textwidth}{}}
\vspace{-2.5\baselineskip}
\caption{Estimated YMC sizes depend strongly on data resolution.
The four datasets being plotted are 
NGC~3351 at 8~pc resolution (quoted in beam FWHM; this work), 
the Antennae galaxies at 12~pc \citep{He_etal_2022}, 
NGC~4945 at 2.2~pc \citep{Emig_etal_2020}, and NGC~253 at 0.5~pc \citep{Levy_etal_2021}.
The dashed and dotted lines mark the identity relation and a factor of three above it.
}
\label{fig:compare_size}
\end{figure}

We use the estimated (sub-)millimeter source sizes as indicators of the YMC candidate physical sizes.
At the distance of NGC~3351, the (beam-deconvolved) half-light radii correspond to 2--8~pc for the 11 sources detected at 350~GHz (Table~\ref{tab:YMC_obs}).
This range broadly overlaps the typical range of radii measured for YMCs in the Milky Way \citep[$R\sbsc{hl}\sim0.5{-}5$~pc;][]{PortegiesZwart_etal_2010} and in other extragalactic systems \citep[$R\sbsc{hl}\sim2{-}5$~pc;][]{Ryon_etal_2017}.

For the YMCs that are only detected at 93~GHz, we instead use their estimated sizes from the 93~GHz image, which spans a similar range (1--6~pc).
To be more precise, the 93~GHz free-free source sizes should reflect the sizes of the \HII\ regions associated with the YMCs rather than the star clusters themselves.
Nonetheless, for the eight sources detected in both bands, we find reasonable agreement between their 93~GHz and 350~GHz sizes (Table~\ref{tab:YMC_obs}), which justifies this choice. 

We note that the YMC size estimates may be affected by systematic effects due to finite data resolution (in addition to the statistical uncertainties quoted in Table~\ref{tab:YMC_obs}).
Our estimated source sizes are all within a factor of three of the beam size, as seen in similar studies for other galaxies over a range of data resolution \citep[Figure~\ref{fig:compare_size}; also see][]{Leroy_etal_2018,Emig_etal_2020,Levy_etal_2021,Levy_etal_2022,He_etal_2022}.
When examining our 350~GHz image at the $0\farcs13\sim6$~pc native resolution, we also find tentative evidence for substructures \textit{within} some sources identified at the $0\farcs17\sim8$~pc working resolution.
These observations suggest that the estimated source sizes may become smaller as one pushes to smaller beam sizes, which was indeed the case with studies of YMCs in NGC~253 \citep{Leroy_etal_2018,Levy_etal_2021}.
That being said, the differences between our 350~GHz images at $0\farcs13$ and $0\farcs17$ are only marginal, so we choose to work with the latter for more straightforward flux and size comparisons with the 93~GHz data.
We expect to test the resolution effect in a follow-up study with ALMA Cycle~9 observations (partially executed, PI: J.~Sun), which will improve the data resolution by a factor of three.

\subsection{Total Mass \& Gas Fraction}
\label{sec:results:Mtot_fgas}

Adding together the stellar and gas mass estimates (or upper limits) for each YMC candidate, we find total masses of $M\sbsc{tot} \equiv M\sbsc{gas}+M_\star \approx 0.3{-}3\times10^5\;\uM$ (Table~\ref{tab:YMC_phys}).
The estimated gas fraction, $f\sbsc{gas} \equiv M\sbsc{gas} / (M\sbsc{gas}+M_\star)$, ranges from $\lesssim$25\% to $\gtrsim$70\%.
This suggests that the YMC candidates identified from our ALMA data span a wide range of evolutionary stages, from the gas mass-dominated early phase to the stellar mass-dominated late phase of cluster formation.

It is worth emphasizing that the total mass and gas fraction estimates can be affected by the ${\sim}0.5$~dex systematic uncertainties on $M\sbsc{gas}$ and $M_\star$ mentioned in Section~\ref{sec:results:Mgas} \& \ref{sec:results:Mstar}.
For example, because $M_\star$ as calculated from Equation~\ref{eq:Mstar} can vary by a factor of 2--5 depending on the assumed stellar age, the $f\sbsc{gas}$ estimates should be similarly sensitive to this assumption.
Indeed, \citet{Leroy_etal_2018} and \citet{Mills_etal_2021} assumed zero-age stellar population for the YMCs in NGC~253 and found comparable stellar and gas masses (modulo synchrotron contamination);
in contrast, \citet{Emig_etal_2020} assumed a 5~Myr age for the YMCs in NGC~4945 and found them to be stellar mass-dominated.
While the different ages assumed in these works are well motivated by known differences of the host galaxies, it is still possible that these different assumptions may partly cause the differing results between studies.
This underlines the importance of accurate YMC stellar age estimates, which we expect to improve with the joint HST+JWST SED fitting.

\subsection{Virial Mass}
\label{sec:results:Mvir}

\begin{figure}[b]
\centering
\gridline{\fig{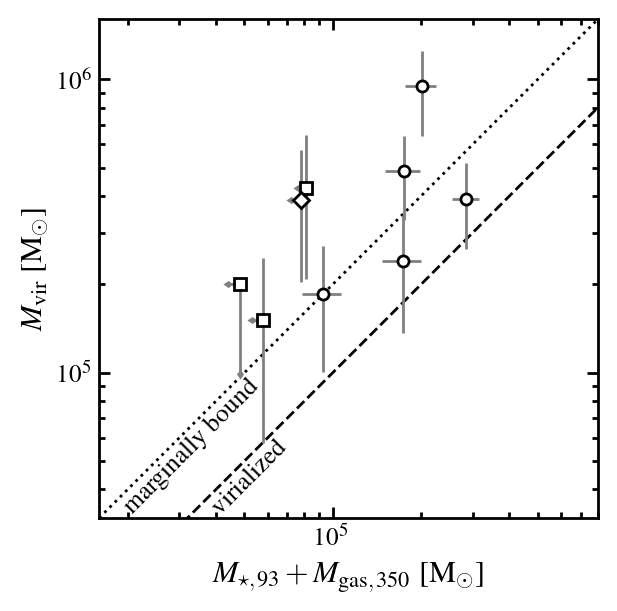}{0.48\textwidth}{}}
\vspace{-2.5\baselineskip}
\caption{YMC total (gas+star) mass versus virial mass, with the latter being calculated from the continuum sizes and \HCN10\ line widths.
The gas reservoirs around most YMCs appear gravitationally unbound or only marginally bound, though the large error budgets on both axes (including statistical and systematic errors; the former shown by the error bars) mean that such inference is inevitably uncertain.}
\label{fig:Mtot_vs_Mvir}
\end{figure}

We use the measured radius ($R\sbsc{hl}$; see Section~\ref{sec:results:size}) and gas velocity dispersion (HCN\;1--0 line width, $\sigma\sbsc{HCN}$; see Section~\ref{sec:analyses:characterize}) to derive the virial mass for each object:
\begin{align}
M\sbsc{vir} 
&\equiv \frac{5 \,\sigma\sbsc{HCN}^2\,R\sbsc{hl}}{f\,G}\nonumber\\
&\approx 7.0\times10^4\,\uM \left(\frac{\sigma\sbsc{HCN}}{10\,\mathrm{km/s}}\right)^2 \left(\frac{R\sbsc{hl}}{1\,\mathrm{pc}}\right) \left(\frac{f}{5/3}\right)^{-1}~.
\label{eq:Mvir}
\end{align}
\noindent Here the geometrical factor $f$ depends on the density profile of the YMCs \citep{Bertoldi_McKee_1992}.
We adopt a density profile of $\rho \propto r^{-2}$ following \citet{Leroy_etal_2018}, which corresponds to $f = 5/3$.

We find virial masses of $10^5{-}10^6\,\uM$ for the 12 YMC candidates with \HCN10\ line measurements available (Table~\ref{tab:YMC_phys}).
Most of them have $M\sbsc{vir} \gtrsim 2\,M\sbsc{tot}$, which seems to imply their gas reservoirs are not gravitationally bound or only marginally bound (Figure~\ref{fig:Mtot_vs_Mvir}).
However, the estimated virial masses have large statistical and systematic uncertainties due to the uncertain size and velocity dispersion measurements, and they also tend to be overestimated because 
(1) the actual YMC sizes may be smaller than those measured at our data resolution (Section~\ref{sec:results:size}), and 
(2) the HCN velocity dispersion may also be biased high due to contamination by diffuse emission from the lower-density ambient gas, despite our best attempt in removing them (Section~\ref{sec:analyses:characterize}).
With these caveats in mind, we conclude that the measured sizes and gas velocity dispersion for the YMC candidates are consistent with them having high total masses of $\gtrsim10^5\,\uM$.

\subsection{Escape Velocity}
\label{sec:results:Vesc}

We estimate the escape velocity for all YMC candidates from their total masses and radii:
\begin{align}
v\sbsc{esc} = \sqrt{\frac{2 G (M\sbsc{tot}/3.4)}{R\sbsc{hl}}}~.
\label{eq:vesc}
\end{align}
\noindent The factor of 3.4 in the numerator accounts for the fraction of mass inside the FWHM of a 3D Gaussian \citep[following][]{Leroy_etal_2018}, which is appropriate since we are using $R\sbsc{hl}$ (i.e., the deconvolved source HWHM; see Section~\ref{sec:analyses:characterize}) in the denominator.

We find escape velocities of $6{-}10\;\uV$ among all YMC candidates (Table~\ref{tab:YMC_phys}).
These estimates are uncertain by ${\sim}0.3$~dex due to systematic errors on the mass estimates; they may also be biased low if $R\sbsc{hl}$ is overestimated at the current data resolution.
Considering these factors, the true escape velocities are likely comparable to, if not larger than, the characteristic sound speed of photoionized gas ($\sim$10$\;\uV$).
In this case, we expect photoionization feedback from massive stars to be not as effective in removing gas and stopping star formation in these dense structures as they would be in less dense structures \citep[consistent with a defining criterion for young massive protoclusters; see][]{Bressert_etal_2012}.
In addition, the estimated $v\sbsc{esc}$ almost certainly exceed the previously estimated threshold of $\sim$1$\;\uV$ for effective gas ejection by protostellar outflows \citep{Matzner_Jumper_2015}, in line with the notion that this particular feedback channel has limited impact on cluster formation \citep[e.g.,][]{Nakamura_Li_2014}.
These arguments imply relatively high star formation efficiency during the birth of these YMCs and the need for other feedback mechanisms to remove the remaining gas \citep[e.g., radiation pressure and stellar winds; see][also see Section~\ref{sec:results:Sigmagas}]{Levy_etal_2021,Menon_etal_2023,Polak_etal_2023}.

\begin{deluxetable*}{lcccccccccccc}[t]
\tablecaption{Physical Properties of the ALMA YMC Candidates\label{tab:YMC_phys}}
\tablewidth{0pt}
\tablehead{
\colhead{ID} &
\colhead{$R\sbsc{hl}$} &
\colhead{$M\sbsc{gas,\,350}$} &
\colhead{$M\sbsc{gas,\,HCN}$} &
\colhead{$M\sbsc{\star,\,93}$} &
\colhead{$M\sbsc{\star,\,HST}$} &
\colhead{$f\sbsc{gas}$} &
\colhead{$M\sbsc{tot}$} &
\colhead{$M\sbsc{vir}$} &
\colhead{$v\sbsc{esc}$} &
\colhead{$t\sbsc{ff}$} &
\colhead{$\Sigma\sbsc{gas}$} \\
\colhead{} &
\colhead{[pc]} &
\colhead{$\log\,[\uM]$} &
\colhead{$\log\,[\uM]$} &
\colhead{$\log\,[\uM]$} &
\colhead{$\log\,[\uM]$} &
\colhead{} &
\colhead{$\log\,[\uM]$} &
\colhead{$\log\,[\uM]$} &
\colhead{[km/s]} &
\colhead{[Myr]} &
\colhead{$[\uSig]$} \\
\colhead{(1)} &
\colhead{(2)} &
\colhead{(3)} &
\colhead{(4)} &
\colhead{(5)} &
\colhead{(6)} &
\colhead{(7)} &
\colhead{(8)} &
\colhead{(9)} &
\colhead{(10)} &
\colhead{(11)} &
\colhead{(12)} \\[-2em]
}
\startdata
\\[-1.5em]
1$^*$ & $<$$4.3$ & $<$$4.3$ & \phs$\cdots$ & \phs$4.4$ & $\cdots$ & $<$$0.40$ & $4.4$--$4.7$ & \phs$\cdots$ & $>$$4.0$ & $<$$1.7$\phn & $<$$240$ \\
2 & \phs$2.9$ & $<$$4.3$ & \phs$4.7$ & \phs$4.6$ & $4.7$ & $<$$0.32$ & $4.6$--$4.8$ & \phs$5.2$ & $5.8$--$7.0$ & $0.64$--$0.78$ & $<$$240$ \\
3 & \phs$2.9$ & \phs$4.8$ & \phs$4.9$ & \phs$4.4$ & $\cdots$ & \phs$0.73$ & $5.0$ & \phs$5.3$ & \phs$8.9$ & \phs$0.51$ & \phs$1710$\phn \\
4$^*$ & $<$$3.3$ & $<$$4.3$ & \phs$\cdots$ & \phs$4.3$ & $\cdots$ & $<$$0.46$ & $4.3$--$4.6$ & \phs$\cdots$ & $>$$4.0$ & $<$$1.3$\phn & $<$$240$ \\
5 & \phs$6.6$ & \phs$5.0$ & $<$$5.1$ & \phs$4.6$ & $\cdots$ & \phs$0.70$ & $5.2$ & \phs$\cdots$ & \phs$7.4$ & \phs$1.4$\phn & \phs$510$ \\
6 & \phs$5.4$ & \phs$5.0$ & \phs$5.4$ & \phs$5.0$ & $5.6$ & \phs$0.51$ & $5.3$ & \phs$6.0$ & \phs$9.7$ & \phs$0.86$ & \phs$760$ \\
7 & \phs$4.1$ & \phs$4.8$ & \phs$5.1$ & \phs$5.0$ & $5.6$ & \phs$0.41$ & $5.2$ & \phs$5.4$ & \phs$10.4$\phn & \phs$0.60$ & \phs$940$ \\
8 & $<$$4.6$ & $<$$4.3$ & \phs$4.8$ & \phs$4.5$ & $5.3$ & $<$$0.40$ & $4.5$--$4.7$ & $<$$5.3$ & $>$$4.0$ & $<$$1.8$\phn & $<$$250$ \\
9 & \phs$4.1$ & \phs$4.7$ & \phs$5.0$ & $<$$4.4$ & $\cdots$ & $>$$0.69$ & $4.7$--$4.9$ & \phs$5.6$ & $5.7$--$6.9$ & $0.92$--$1.1$\phn & \phs$690$ \\
10 & \phs$2.6$ & \phs$4.6$ & $<$$4.7$ & $<$$4.4$ & $\cdots$ & $>$$0.64$ & $4.6$--$4.8$ & \phs$\cdots$ & $6.6$--$8.2$ & $0.48$--$0.60$ & \phs$1480$\phn \\
11 & \phs$4.0$ & \phs$4.8$ & $<$$4.8$ & $<$$4.4$ & $\cdots$ & $>$$0.70$ & $4.8$--$4.9$ & \phs$\cdots$ & $6.1$--$7.2$ & $0.85$--$1.0$\phn & \phs$800$ \\
12 & \phs$6.2$ & \phs$5.2$ & \phs$5.2$ & \phs$4.4$ & $\cdots$ & \phs$0.85$ & $5.2$ & \phs$5.7$ & \phs$8.4$ & \phs$1.1$\phn & \phs$850$ \\
13$^*$ & $<$$4.5$ & $<$$4.3$ & \phs$\cdots$ & \phs$4.5$ & $\cdots$ & $<$$0.40$ & $4.5$--$4.7$ & \phs$\cdots$ & $>$$4.1$ & $<$$1.7$\phn & $<$$260$ \\
14 & \phs$3.3$ & \phs$4.5$ & $<$$4.8$ & \phs$4.6$ & $\cdots$ & \phs$0.45$ & $4.9$ & \phs$\cdots$ & \phs$7.5$ & \phs$0.68$ & \phs$670$ \\
15 & \phs$7.6$ & \phs$5.2$ & \phs$5.3$ & \phs$5.0$ & $\cdots$ & \phs$0.61$ & $5.5$ & \phs$5.6$ & \phs$9.8$ & \phs$1.2$\phn & \phs$670$ \\
16$^*$ & $<$$3.4$ & $<$$4.3$ & \phs$\cdots$ & \phs$4.3$ & $\cdots$ & $<$$0.51$ & $4.3$--$4.6$ & \phs$\cdots$ & $>$$3.8$ & $<$$1.4$\phn & $<$$270$ \\
17 & \phs$4.2$ & $<$$4.3$ & \phs$5.0$ & \phs$4.8$ & $5.3$ & $<$$0.24$ & $4.8$--$4.9$ & \phs$5.6$ & $6.0$--$7.0$ & $0.94$--$1.1$\phn & $<$$260$ \\
18 & \phs$5.3$ & \phs$4.9$ & $<$$4.9$ & \phs$5.1$ & $5.4$ & \phs$0.41$ & $5.3$ & \phs$\cdots$ & \phs$9.6$ & \phs$0.84$ & \phs$630$ \\
\enddata
\tablecomments{
(2) half-light radius (or upper limit in case of unsuccessful beam deconvolution; \S\ref{sec:results:size});
(3) gas mass based on 350~GHz continuum detections and upper limits (\S\ref{sec:results:Mgas});
(4) gas mass based on \HCN10\ line detections and upper limits (\S\ref{sec:analyses:characterize}, \S\ref{sec:results:Mgas});
(5) stellar mass based on 93~GHz continuum detections and upper limits (\S\ref{sec:results:Mstar});
(6) stellar mass from UV--optical SED fitting for sources with cross-matched HST clusters (\S\ref{sec:analyses:crossmatch});
(7) gas mass fraction (or upper/lower limit for non-detection at 350/93~GHz; \S\ref{sec:results:Mtot_fgas});
(8) total mass (or 5$\sigma$ range for non-detection in either band; \S\ref{sec:results:Mtot_fgas});
(9) virial mass (for those with measured gas velocity dispersion, \S\ref{sec:results:Mvir});
(10) escape velocity (or 5$\sigma$ range / lower limit according to columns 2 \& 8; \S\ref{sec:results:Vesc});
(11) free-fall time (or 5$\sigma$ range / upper limit according to columns 2 \& 8; \S\ref{sec:results:rho_tff});
(12) gas surface density (or upper limit based on 350~GHz continuum sensitivity, \S\ref{sec:results:Sigmagas}).
We note that most quantities reported here have large systematic uncertainties.
The YMC radii are likely affected by finite data resolution (\S\ref{sec:results:size}); the gas and stellar mass estimates have ${\sim}0.5$~dex systematic error from the assumptions involved in their derivations (\S\ref{sec:results:Mgas}--\ref{sec:results:Mstar}); other columns are also affected by the propagation of these uncertainties (\S\ref{sec:results:Mtot_fgas}--\ref{sec:results:Sigmagas}).}
\tablenotetext{*}{These four sources are detected only in 93~GHz continuum. They may be supernova remnants with strong synchrotron emission rather than YMCs (see \S\ref{sec:discuss:nature}), in which case the $M_{\star,\,93}$ values and subsequent calculations would not be reliable.}
\vspace{-1\baselineskip}
\end{deluxetable*}

\vspace{-2\baselineskip}

\subsection{Volume Density \& Free-fall Time}
\label{sec:results:rho_tff}

We also derive the volume density and gravitational free-fall time from the YMC total masses and radii:
\begin{align}
\rho\sbsc{tot} &= \frac{M\sbsc{tot}/3.4}{\frac{4}{3} \pi R\sbsc{hl}^3}~, \label{eq:rho}\\
t\sbsc{ff} &= \sqrt{\frac{3\pi}{32 G \rho\sbsc{tot}}}~. \label{eq:tff}
\end{align}
\noindent We similarly include a factor of 3.4 in Equation~\ref{eq:rho} as in Equation~\ref{eq:vesc}.
That is, we are calculating the mean mass volume density within a sphere defined by the (deconvolved) FWHM of each source, and then the free-fall time corresponding to that volume density.

We find volume densities of $50{-}200\,\urho$, which is equivalent to hydrogen nuclei number densities of $1.5{-}6\times10^3\,\mathrm{cm^{-3}}$.
The corresponding free-fall times are $0.5{-}1$~Myr (Table~\ref{tab:YMC_phys}), with ${\sim}0.3$~dex of systematic uncertainty coming from the mass estimates and a potential bias towards larger values due to marginally resolved sizes (see Section~\ref{sec:results:size}).
Given that the YMC formation process can last a couple of free-fall times \citep{Skinner_Ostriker_2015}, it is possible that these YMCs are still forming and possess a substantial gas reservoir at ages of $\sim$3~Myr (also see related discussions in Sections~\ref{sec:results:Vesc}--\ref{sec:results:Sigmagas}).
Moreover, the short free-fall time relative to the typical supernova explosion delay time ($\gtrsim$3~Myr) means that the initial gravitational collapse may be too fast for supernova feedback to play a role \citep{Fall_etal_2010}.
This inference again supports fairly high star formation efficiencies when forming these dense YMCs.

\subsection{Gas Surface Density}
\label{sec:results:Sigmagas}

We calculate the gas surface density near the center of each YMC candidate via:
\begin{align}
\Sigma\sbsc{gas} = \frac{M\sbsc{gas}}{A\sbsc{eff}} = \frac{M\sbsc{gas}}{\pi R\sbsc{hl}^2/\ln{2}}~.
\label{eq:Sigmagas}
\end{align}
\noindent Here $A\sbsc{eff}=2\pi\sigma\sbsc{xy}^2=\pi R\sbsc{hl}^2/\ln{2}$ is the effective area of the 2D Gaussian fit for each detected YMC in 350~GHz (Section~\ref{sec:analyses:characterize}). The second equality follows from the conversion between the Gaussian HWHM (i.e., $R\sbsc{hl}$) and dispersion: $\sigma\sbsc{xy}=R\sbsc{hl}/\sqrt{2\ln{2}}$.

We find gas surface densities of $500{-}2000\,\uSig$ for the YMC candidates with 350~GHz continuum detections (Table~\ref{tab:YMC_phys}), with a ${\sim}0.5$~dex systematic uncertainty associated with $M\sbsc{gas}$ and a potential bias towards lower values due to possibly overestimated sizes at current data resolution.
The estimated as surface densities are much higher than typical values of giant molecular clouds \citep[${\sim}10^2\,\uSig$;][]{Heyer_Dame_2015,Rosolowsky_etal_2021,Sun_etal_2022} and also higher than the median value for dense clumps in the Milky Way \citep[${\sim}500\;\uSig$;][]{Urquhart_etal_2018}.
However, they are close to surface density thresholds above which one expects high star formation efficiency, as shown by analytical and numerical studies \citep[${\gtrsim}\,10^3\;\uSig$; see e.g.,][]{Fall_etal_2010,KimJG_etal_2018,Menon_etal_2023}.

The gas surface density can play a key role in regulating the star formation efficiency because it determines the effectiveness of various forms of feedback in destroying the natal molecular clouds \citep[e.g., see section~5 of][]{Chevance_etal_2022a}.
In particular, radiation pressure is believed to be the most important feedback process at high surface densities.
For direct (UV) radiation pressure, one can derive an Eddington ratio, or the ratio of radiation pressure force to gravitational force, from the gas surface density and gas mass fraction 
\begin{align}
f\sbsc{Edd} \approx \frac{4\Psi\,(1-f\sbsc{gas})}{3\pi G c\,\Sigma\sbsc{gas}\,(4-3f\sbsc{gas})}~,
\label{eq:fEdd}
\end{align}
\noindent where $\Psi\approx10^3\,L_\odot/M_\odot$ is the light-to-mass ratio of a central stellar population with a Kroupa IMF and an age of ${\lesssim}\,3$~Myr (based on a Starburst99 simulation).
The above is derived for a uniform gas sphere with a central star cluster, but other spatial distributions give similar results \citep[e.g.,][]{Raskutti_etal_2017,Reissl_etal_2018,Krumholz_etal_2019}.
From this equation, we estimate $f\sbsc{Edd}\sim0.2{-}0.7$ for the YMC candidates with associated gas, which implies that their direct radiation pressure may not (yet) be enough to expel the gas \citep[assuming the gas column density variation from sightline to sightline is small; see][]{Thompson_Krumholz_2016,Raskutti_etal_2016,Raskutti_etal_2017}.

Furthermore, at $\Sigma\sbsc{gas}\gtrsim10^3\,\uSig$, the star formation efficiency is expected to exceed $50\%$ when limited only by direct FUV radiation pressure on dust in combination with the rocket effect and photoevaporation from EUV \citep[][]{KimJG_etal_2018,HeCC_etal_2019}. 
Stellar winds, since they have a similar momentum flux to that from radiation, would not significantly change the expected star formation efficiency \citep{Lancaster_etal_2021,Polak_etal_2023}.
Unless the mass function is top-heavy or the dust abundance is enhanced, reprocessed infrared radiation has $f\sbsc{Edd} = \kappa\sbsc{IR} \Psi/(4 \pi G c) < 1$ and therefore does not aid in limiting star formation either \citep{Skinner_Ostriker_2015,Menon_etal_2022}.
To conclude, the high gas surface densities found for many of our YMC candidates suggest that the remaining gas may continue to feed star formation, leading to an overall high star formation efficiency.

Other than the implications on feedback and star formation efficiency, the measured gas surface densities also imply $V$-band extinctions of $\sim$30--110~mag \citep[assuming Galactic dust abundance and extinction curve;][]{Draine_2011}, or $\sim$15--55~mag if half of the gas is in front of the stellar body.
Such high extinction helps explain the lack of optical (or even IR) counterparts for many of our ALMA sources, as was found for YMCs in other systems \citep[e.g.,][]{Leroy_etal_2018,He_etal_2022}.


\section{Discussion} \label{sec:discuss}

In Section~\ref{sec:results}, we report the measured physical properties for the 18 YMC candidates identified from our new ALMA data and complemented by JWST and HST data.
Our measurements suggest that most of these objects are massive, compact, in the early phase of formation, and likely associated with high star formation efficiency.
Here we build on the quantitative results and address a few key remaining questions:
(1) What do the rich multiwavelength measurements tell us about the nature and evolutionary stages of the YMC candidates?
(2) How do the ALMA-identified YMCs relate to the exposed, more evolved clusters?
(3) How do the YMCs fit in the large-scale context of the central starburst ring in NGC~3351?

\subsection{Nature and Evolutionary Stages of the ALMA-identified YMC Candidates}
\label{sec:discuss:nature}

Based on the multiwavelength observational properties of the 18 ALMA sources and their cross-matching results with the JWST and HST data, we can classify most of them into four categories. These categories are inspired by the classification schemes for star clusters introduced in \citet{Johnson_2005} and \citet{Whitmore_etal_2014} as well as a similar scheme for molecular clouds used in \citet{Kawamura_etal_2009}.

\begin{enumerate}[leftmargin=1em,itemsep=0.1em,topsep=0.4em]

\item[$\bullet$] \textit{Type 1: Starless Clumps (\#9, \#10, \#11).}
These objects have substantial gas reservoirs indicated by thermal dust emission at 350~GHz, yet they show no signs of star formation through UV--optical stellar photospheric emission or 93~GHz free-free emission from \HII\ regions.
They are likely dense gas clumps on their way to becoming YMCs, representing the earliest phase of YMC formation.

\item[$\bullet$] \textit{Type 2: Clump--\HII\ region complexes (\#3, \#5, \#12):}
These objects have both gas reservoirs producing dust emission at 350~GHz and associated \HII\ regions producing free-free emission at 93~GHz, but the stellar content remains largely invisible in UV-to-IR bands due to high extinction.
They represent the deeply embedded phase of YMC formation, when a substantial stellar body (including massive stars) has formed but the gas reservoir is neither expelled nor exhausted.

\item[$\bullet$] \textit{Type 3: Clump--\HII\ region--cluster complexes (\#6, \#7, \#14, \#15, \#18):}
These objects are detected simultaneously in 350~GHz dust emission, 93~GHz free-free emission, as well as PAH and stellar photospheric emission in the near- to mid-IR;
three out of five are also visible in optical bands.
These are exposed but still forming clusters as they still have ample gas left and \HII\ regions glowing around massive stars.

\item[$\bullet$] \textit{Type 4: Exposed \HII\ region--cluster complexes (\#2, \#8, \#17):}
These objects no longer have detectable 350~GHz emission, suggesting that much of the local gas reservoir has been expelled or exhausted.
They are still visible in both 93~GHz free-free emission and UV-to-IR stellar photospheric emission.
These are emerging young clusters that have stopped forming and likely survived the violent gas expulsion process near the end of their formation process.

\end{enumerate}

There are four ALMA sources (\#1, \#4, \#13, and \#16) that do not belong in any of the above categories.
They are visible at 93~GHz but almost completely missing in all other wavelengths. 
Considering the stringent upper limits we can put on the gas surface density based on non-detections of the 350~GHz continuum, the \HCN10\ line, and even the \CO32\ line (see Appendix~\ref{apdx:spectra}), these objects cannot be deeply embedded and should be visible at least in the near-IR if they are indeed forming YMCs with \HII\ regions glowing intensely in free-free emission.

One possible explanation is that these 93~GHz-only sources are not forming YMCs, but rather supernovae remnants (SNRs) with their 93~GHz emission dominated by synchrotron radiation.
As mentioned in Section~\ref{sec:results:Mstar}, previous 1.4~GHz MERLIN observations \citep{Hagele_etal_2010} put an upper limit of $0.3\;\uS$ for any source more compact than the $\sim$10~pc MERLIN beam.
Combining this constraint at 1.4~GHz with the measured 93~GHz flux density of $0.03{-}0.04\;\uS$ for the four ALMA sources in qustion, the implied radio spectral index is ${\gtrsim-}0.5$.
This is consistent with a synchrotron spectrum that turns over at an intermediate frequency due to synchrotron self-absorption, which is very probable given the small sizes and strong magnetic field of SNRs (e.g., see \citealt{Lenc_Tingay_2009} for spectra of SNRs in NGC~4945 with turnover frequencies of ${\sim}5$~GHz).
Furthermore, \citet{Hagele_etal_2010} showed that a SNR at their detection threshold of $0.3\;\uS$ is expected to have a diameter of $\sim$2~pc.
SNRs slightly fainter and larger in size would remain undetected in the MERLIN observations, while they can be detected but unresolved in our ALMA observations, just like the four 93~GHz-only sources.
We thus suggest that these sources may actually be SNRs with synchrontron emission dominating the 93~GHz continuum.
Future observations with the Karl~G.~Jansky Very Large Array (VLA) at 3--30~GHz with matched resolution can help verify this hypothesis.

\subsection{YMCs versus Evolved Clusters}
\label{sec:discuss:timescale}

\begin{figure*}[t]
\centering
\gridline{
\fig{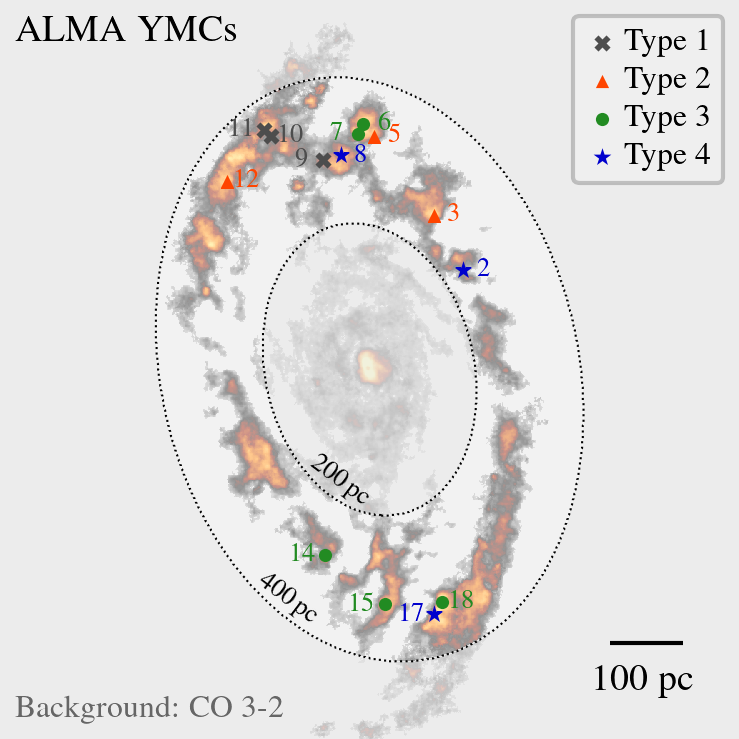}{0.49\textwidth}{}
\hfill
\fig{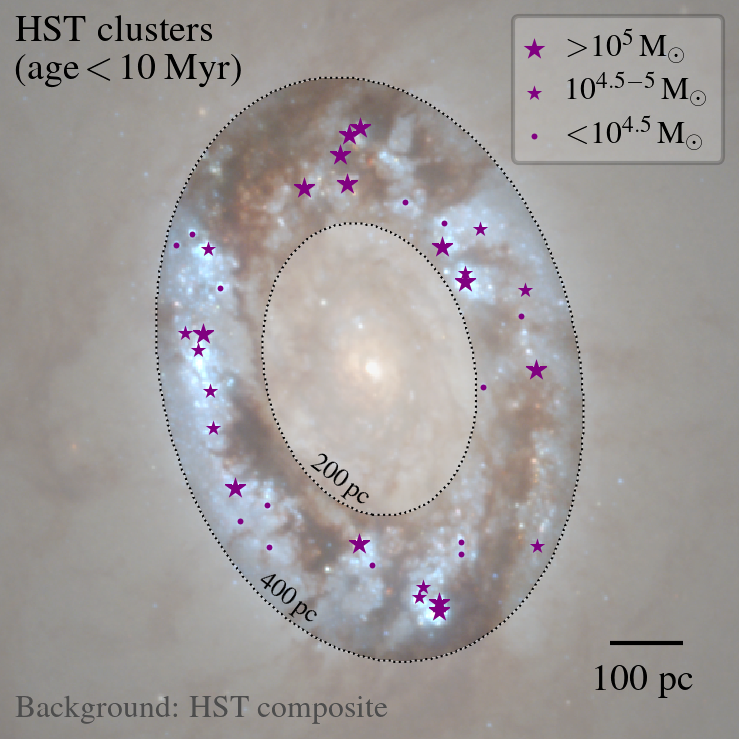}{0.49\textwidth}{}
}
\vspace{-2.0\baselineskip}
\caption{
\textit{Left:} ALMA-identified YMCs labeled according to their assigned categories in  Section~\ref{sec:discuss:nature}.
The background \CO32\ image is shaded in a way to highlight structures along the starburst ring ($r\sbsc{gal}=200{-}400$~pc).
\textit{Right:} HST-identified young clusters (age$\;<\;$10~Myr) labeled according to their estimated stellar mass, with the background image similarly shaded.
These young clusters (especially the more massive ones) are likely the direct descendants of the ALMA-identified YMCs.
}
\vspace{0.5\baselineskip}
\label{fig:young_old}
\end{figure*}

\begin{figure*}[hbt]
\centering
\gridline{\fig{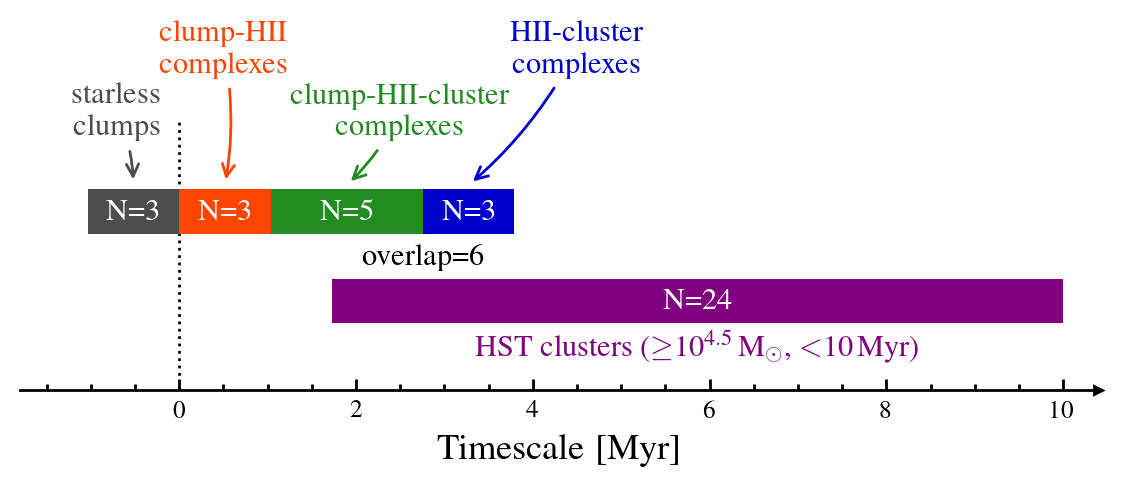}{0.7\textwidth}{}}
\vspace{-2.0\baselineskip}
\gridline{\fig{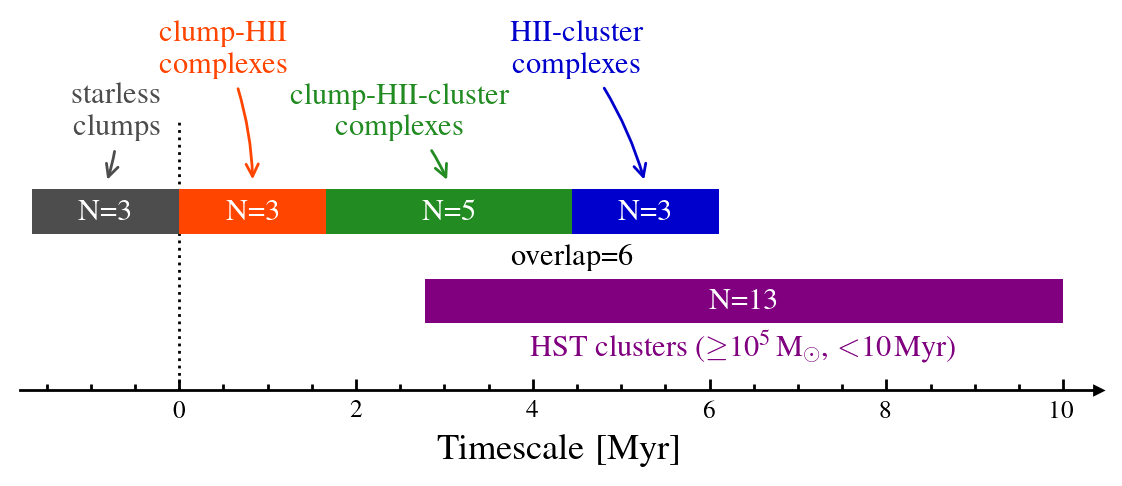}{0.7\textwidth}{}}
\vspace{-2.0\baselineskip}
\caption{
Two plausible ways of inferring the YMC formation timeline by matching ALMA YMCs (Types~1--4) with young HST clusters (age$\,<\,$10~Myr).
\textit{Top:} Estimated timeline using HST clusters $\geq$~$10^{4.5}\,\uM$ as the reference sample, which matches the nominal mass range probed by the ALMA continuum observations.
The four phases of YMC formation observable with ALMA are inferred to last for 1.0--1.0--1.8--1.0~Myr, respectively.
\textit{Bottom:} Estimated timeline using HST clusters $\geq$~$10^{5}\,\uM$ as the reference sample, which accounts for a ${\sim}0.5$~dex median offset between the 93~GHz-based and optical SED-based stellar mass estimates (see \S\ref{sec:results:Mstar}).
The inferred durations of the four phases are 1.7--1.7--2.8--1.7~Myr in this case.
}
\vspace{0.5\baselineskip}
\label{fig:timeline}
\end{figure*}

In the previous section, we classified YMCs into four categories, which likely correspond to four evolutionary stages in the cluster formation process.
However, there is a fifth stage, i.e., evolved clusters without associated gas or \HII\ regions, that has been omitted from the discussion above.
This omission is obviously due to our target selection based on ALMA data -- older clusters without 93~GHz free-free or 350~GHz dust emission should not be in our list of ALMA-identified sources to begin with.
These older clusters can nonetheless be identified with the HST and JWST data; they are also of great interest as they represent the end products of the cluster formation process.
Here, we explicitly connect the ALMA-identified YMCs to more evolved clusters included in the HST cluster catalog (Section~\ref{sec:data:HST}) to fill in the last missing piece.

One important consideration for such cross-dataset comparison is the difference in data sensitivity, which maps into different mass sensitivity for cluster detection.
The ALMA continuum observations were designed to probe forming YMCs down to $\sim10^{4.5}\,\uM$ in both gas and stellar mass.
The HST broad-band observations, in contrast, can detect exposed clusters with much lower masses \citep[$\sim10^{3.5}\,\uM$;][]{Lee_etal_2022}.
In order to compare the cluster populations detectable by these datasets in a sensible way, we choose to focus on HST clusters above some matching mass threshold.
Given the systematic uncertainties in the mass estimates from both sides, we use two thresholds to bracket a sensible range: one at $10^{4.5}\,\uM$ to directly match the nominal sensitivity limit of the ALMA data, the other at $10^{5}\,\uM$ to account for the median ${\sim}0.5$~dex offsets between the 93~GHz-based and UV--optical SED-based stellar mass estimates for the ALMA-HST cross-matched sources (see Table~\ref{tab:YMC_phys}).

The HST cluster catalog includes in total 42 star clusters in the central region of NGC~3351 ($r\sbsc{gal}$\,$<$\,500~pc).
The vast majority (37) of them have SED-based ages$\,<\,$10~Myr and are located between $r\sbsc{gal}=200{-}400$~pc (Figure~\ref{fig:young_old} right panel).
The same $r\sbsc{gal}$ range covers all 14 ALMA-identified YMCs (including the six ALMA-HST cross-matches and excluding the four sources that are possibly SNRs) and almost all gas structures associated with the starburst ring.
Since the orbital period along this ring is also $\sim$10~Myr \citep{Rubin_etal_1975,Hagele_etal_2007,Leaman_etal_2019}, we do not expect much radial migration for these $<$10~Myr clusters.
We thus view these young HST clusters along the ring as either counterparts or direct descendants of the ALMA YMCs (modulo caveats with mass range matching).

From the number counts of ALMA YMCs and young HST clusters within the matched mass range, we can further infer the timescales of the various evolutionary stages during YMC formation \citep[similar to the method used for determining molecular cloud evolution timescales by][]{Kawamura_etal_2009}.
Here we assume that
(1) the SFR of the starburst ring does not change drastically over a 10~Myr timescale;
(2) the combined ALMA+HST YMC sample is complete over the matched mass range up to an age of ${\sim}10$~Myr;
and (3) there is no substantial loss of YMCs over this period.
Under these assumptions, the number count of YMCs in each evolutionary stage should be proportional to the average duration of that stage.
Furthermore, the combined sample of ALMA YMCs with observable signs of active star formation (i.e., Types~2--4 with free-free emission) plus HST clusters with age ${\leq}10$~Myr should span a full 10~Myr window from the onset of star formation.
This latter inference provides a necessary reference timescale for converting the relative duration of all stages to absolute values in Myr unit.

We show our inferred evolutionary timeline of YMC formation in Figure~\ref{fig:timeline}.
Depending on the mass range of HST clusters ($\geq$$10^{4.5}\,\uM$ or $\geq$$10^{5}\,\uM$) used as the reference sample, we find that the typical duration of the four stages (mapping to Types~1--4) are 1--1.7~Myr, 1--1.7~Myr, 1.8--2.8~Myr, and 1--1.7~Myr, respectively.
That is, a starless clump lasts $\sim$1--2~Myr (or  ${\sim}2\,t\sbsc{ff}$) before forming the first massive stars capable of creating \HII\ regions.
From that point on, it takes $\sim$1--2~Myr for the star cluster to become visible in the near-IR and $\sim$2--3~Myr to become visible in the optical.
The cold gas reservoir disappears over $\sim$3--4~Myr (or 4--6 $t\sbsc{ff}$), and the associated \HII\ region fades away in $\sim$4--6~Myr.
Overall, these timescales agree well with previous observations and simulations of various types of star-forming regions giving birth to star clusters \citep[e.g.,][]{Whitmore_Zhang_2002,Tan_etal_2006,Reggiani_etal_2011,Whitmore_etal_2014,Whitmore_etal_2023,Skinner_Ostriker_2015,Grasha_etal_2018,Grasha_etal_2019,KimJG_etal_2018,KimJG_etal_2019,Hannon_etal_2019,Hannon_etal_2022,Kruijssen_etal_2019b, LiH_etal_2019,Chevance_etal_2020a,Grudic_etal_2021a,KimJY_etal_2021,KimJY_etal_2023}.

We would like the emphasize that there are substantial uncertainties on the inferred timescales, such that they should be viewed as preliminary, order-of-magnitude estimates.
For example, it is possible that some of the assumptions mentioned above are not appropriate for the starburst ring in NGC~3351.
The 10~Myr orbital period of this system means that the SFR can fluctuate moderately on a similar timescale (see observational constraints by \citealt{Calzetti_etal_2021}; also see demonstrations of such behavior in simulations by \citealt{Armillotta_etal_2019,Sormani_etal_2020,Moon_etal_2022}).
Star clusters may also get destroyed within 10~Myr due to either violent gas removal at the end of their formation process or mass loss due to stellar evolution \citep[see][for a thorough review]{Krumholz_etal_2019}.
These concerns could be addressed, in theory, by choosing even younger HST clusters (e.g., ${<}5$~Myr) as the reference sample, but various sources of systematic effects on the cluster age estimates especially at ${\lesssim}10$~Myr (Section~\ref{sec:data:HST}) would render such analysis unreliable in practice.
On this front, follow-up studies probing more diverse environments (e.g., those with longer dynamical timescales) will be crucial to verify our results in different physical conditions and to improve the currently limited statistics.

The reliability of the inferred timeline is also fundamentally tied to the reliability of the HST cluster measurements themselves.
As discussed in Section~\ref{sec:data:HST}, the revised SED fitting scheme presented in \citet{Thilker_etal_2024} reflects the latest and by-far the most systematic efforts in dealing with photometric degeneracies, but we expect follow-up studies to further refine the fitting results and address remaining issues (especially with the youngest clusters).
For example, these goals can be achieved by employing HST H$\alpha$ narrowband and JWST IR data to complement the HST broadband photometry (e.g., K.~Henny et al., in preparation).
Additional narrowband Pa$\alpha$ imaging with JWST in Cycle~3 (PI: A~Leroy) and/or radio continuum observations with ALMA and VLA in the future would also provide unique constraints to help pin down the ages.

\subsection{YMCs in Large-scale Context}
\label{sec:discuss:context}

After examining the plausible evolutionary phases and timescales of the ALMA YMCs, we now put their properties in the context of the large-scale environment, i.e., the central starburst ring in NGC~3351.

\vspace{0.4\baselineskip}
\noindent$\bullet$ \textit{Star formation rate contributed by the YMCs:} 
The central region in NGC~3351 is long known to be intensely star-forming, with an estimated SFR surface density of $\sim$0.5--0.8$\;\uSigSFR$ and a total SFR of $\sim$0.2--0.5$\;\uSFR$ \citep{Elmegreen_etal_1997,Planesas_etal_1997,Ma_etal_2018,Calzetti_etal_2021,Song_etal_2021}.
In such an extreme condition, the cluster formation efficiency is expected to be high, with 30--60\% of stellar mass formed in clusters \citep[e.g.,][though see \citealt{Chandar_etal_2017,Chandar_etal_2023a} for counter evidence]{Goddard_etal_2010,Adamo_etal_2011,LiH_etal_2018,Mills_etal_2021,Grudic_etal_2022}.

We can use the estimated masses and timescales for the ALMA-identified YMCs to estimate the fractional SFR contributed by these sources alone.
The total stellar mass of all YMCs (excluding the four sources that are possibly SNRs) is $7\times10^5\,\uM$, and the inferred duration of the 93~GHz-bright phase is 4--6~Myr (Section~\ref{sec:discuss:timescale}).
Dividing these two numbers gives us an instantaneous SFR of 0.1--0.2~$\uSFR$, which is already $\sim$50\% of the total SFR of the starburst ring.
Such a high fraction is consistent with similar estimates for the central starburst regions in NGC~253 and NGC~4945 \citep{Leroy_etal_2018,Emig_etal_2020}.
Note that this simple calculation ignores the considerable gas mass associated with the YMCs, and a fraction of that gas may be converted into stars in the future.
Nor does it include the contribution from less massive clusters below our detection threshold.
Given these omissions, the actual fraction of stars formed in clusters may be even higher.

\begin{figure*}[hbt]
\gridline{
\fig{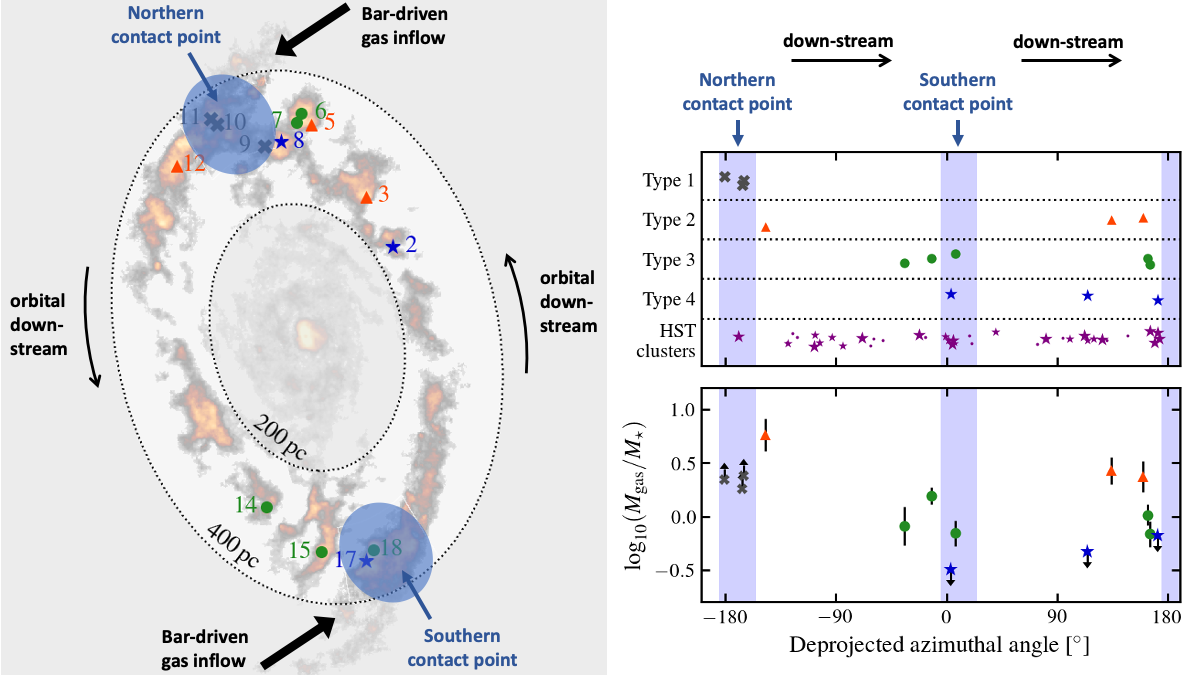}{\textwidth}{}
}
\vspace{-2\baselineskip}
\caption{
\textit{Left:} Key structural features around the starburst ring, including two streams of bar-driven gas inflow from large radii and the two ``contact points'' where the inflow gas collides with the ring material.
Markers and background image are the same as the left panel in Figure~\ref{fig:young_old}.
\textit{Right:} Distributions and properties of the ALMA YMCs and HST clusters, as a function of the deprojected azimuthal angle (increasing counter-clockwise, with zero-point defined by the galaxy position angle).
The blue shaded regions indicate azimuthal angle ranges for the Northern/Southern contact points highlighted in the left panel.
There is tentative evidence for a progression in YMC evolutionary stage (from Type~1 to Type~4) and gradual decrease in YMC gas to stellar mass ratio along the Eastern side of the ring orbit (i.e., from -180$^\circ$ to 0$^\circ$ in azimuthal angle), but the trends are less obvious for the other side.
No clear correspondence or variation is seen in the azimuthal distribution of the HST clusters either.
}
\vspace{0.5\baselineskip}
\label{fig:azimuthal}
\end{figure*}

\vspace{0.4\baselineskip}
\noindent$\bullet$ \textit{Testing ring star formation models with the YMCs:} 
Another noteworthy feature of NGC3351's starburst ring is its clean orbital configuration at a favorable viewing angle (see Figure~\ref{fig:azimuthal} left panel), which stands out among all systems whose YMC populations have been studied by ALMA so far.
As an iconic bar-fed nuclear ring with unambiguous orbital streamlines, this system offers an opportunity to test existing models of star formation in similar systems.

There are at least two influential theoretical models of star-forming rings in the literature \citep[see][for a nice summary]{Boker_etal_2008}.
A first scenario, dubbed the ``popcorn model'', considers star formation triggered by gravitational instability as gas accumulates on the ring and reaches a critical density \citep{Elmegreen_1994}.
As star formation happens stochastically across the system, there may not be any preferable locations along the ring or systematic azimuthal trends in the properties of the young star clusters.

A second scenario, namely the ``pearls on a string model'' \citep{Boker_etal_2008}, suggests that star formation is mostly triggered near or slightly downstream from the contact points, where the gas inflow enters the ring.
An alternative model proposed for the Galactic Central Molecular Zone (CMZ) argues that star formation should be triggered by gas tidal compression near the pericenters of the ring orbit \citep[e.g.,][also see \citealt{Callanan_etal_2021} for similar analyses on M83's center]{Longmore_etal_2013}.
In either case, an evolutionary sequence is expected downstream from the triggering positions, with progressively older clusters further down the orbit.

As the YMCs identified in our ALMA observations probably have ages younger than the orbital period (see Section~\ref{sec:discuss:timescale}), their distribution and azimuthal trends may be particularly helpful for differentiating the competing scenarios described above.
We find that the ALMA YMCs appear to concentrate near the two contact points (Figure~\ref{fig:azimuthal} left), though with a slight preference towards their upstream side.
When examining trends with the deprojected azimuthal angle $\phi$ (Figure~\ref{fig:azimuthal} right), we \textit{do not} see a clear, coherent trend when considering all YMCs along the ring.
There may be tentative evidence for a progression in YMC evolutionary stages from  $\phi=-180^\circ$ to 0$^\circ$ (i.e., the Eastern side of the ring), with Type~1 sources located right at the Northern contact point and subsequent types further down the ring orbit; a similar trend could be argued for the YMC gas to stellar mass ratio.
Nonetheless, these trends become less clear on the other (Western) side of the ring.
There is no clear trend or preferential distribution in the HST clusters either.

A subset of our observational results, especially the concentration of YMCs near the contact points and the possible azimuthal trends, seems to favor the ``pearls on a string'' model.
Notably, the apparent progression from Type~1 to Type~4 YMCs spans half of the ring orbit from the Northern to the Southern contact point, and our estimated 4--5~Myr timescale for such progression (Section~\ref{sec:discuss:timescale}) does agree with the $\sim$5~Myr orbital time across half of the ring \citep{Leaman_etal_2019}.
Nonetheless, these trends are only based on a small number ($\sim$8) of ALMA YMCs and do not show up consistently across the entire ring or in the (more evolved) HST clusters.

We can partly make sense of the somewhat mixed results in light of recent numerical studies of star-forming rings.
For example, \citet{Seo_Kim_2013} showed that the presence or absence of a cluster age sequence may depend on the total SFR of the ring.
\citet{Sormani_etal_2020} found that the instantaneous distribution of clusters can vary substantially with time due to stochasticity of star formation along the ring, and that one may only see clear azimuthal age trends either through time-averaging or when focusing on the youngest clusters ($\leq$0.25~Myr in their system with $\sim$5~Myr orbital period).
\citet{Moon_etal_2021,Moon_etal_2022} showed that local concentration of young clusters upstream from contact points can happen following a temporary, asymmetric boost of gas inflow rate.
Such asymmetry and non-steadiness can naturally arise from clumpiness in bar-driven inflows as well as varying accretion efficiency onto the ring \citep{Sormani_Barnes_2019,Hatchfield_etal_2021}.
Together, these studies highlighted key aspects of star-forming rings that are not entirely captured by the simple ``popcorn'' or the ``pearls on a string'' models, but can help explain some of the observed behaviors in NGC~3351's central ring and its YMC population.

We anticipate more sensitive observations in the future (including partially executed ALMA Cycle~9 observations) to improve the statistics especially for the youngest clusters and allow for more detailed comparisons with simulations.
Studying a large sample of star-forming ring systems (with favorable viewing angles) will be critical for differentiating time-varying effects from persistent trends.

\vspace{0.4\baselineskip}
\noindent$\bullet$ \textit{YMCs as potential drivers of large-scale gas outflows:} 
Central starburst rings are believed to be powerful drivers of multi-phase galactic winds and outflows \citep[e.g.,][]{Armillotta_etal_2019,Nguyen_Thompson_2022}, and there are indications of such outflows from NGC~3351's central ring from existing multiwavelength observations.
By analyzing Chandra X-ray imaging data, \citet{Swartz_etal_2006} found evidence of hot gas expanding beyond the starburst ring and likely above the plane of the galaxy.
This hot gas is estimated to contain thermal energy of ${\sim}f^{1/2}10^{54}\,$erg (here $f\gtrsim10^{-4}$ is the volume filling factor of the gas).
A more recent study based on VLT/MUSE data \citep{Leaman_etal_2019} identified and modelled a stream of warm ionized gas outflow, with a radial velocity of ${\sim}70\,\uV$ and kinetic energy of ${\sim}6\times10^{52}\,$erg.
Both studies highlighted a dust lane visible in optical images to the southeast of the ring (see Figure~\ref{fig:wide_field} left panel, right outside the white box) and interpreted it as a gas shell confining the outflow.
The new JWST/MIRI images (Figure~\ref{fig:wide_field} right panel) clearly show that a similarly shaped gas shell exists on the opposite side, even though it is less obvious from the optical image.
Therefore, gas outflow may be present on all sides of the ring and contain more kinetic energy than estimated before.

The YMCs studied in this work are likely too young to be the main energy source for driving these hot and warm gas outflows.
With our inferred timeline of $\lesssim$4~Myr after birth, most of them have not produced many supernovae to heat the hot X-ray emitting gas; their arrival was also too late to accelerate the aforementioned gas shell to its current location \citep[which was estimated to take ${\gtrsim}10$~Myr by][]{Leaman_etal_2019}.
Besides, we do not see clear evidence of localized outflow around the YMCs in their molecular line spectra either (see Appendix~\ref{apdx:spectra}), unlike those found for some of the YMCs in NGC~253 \citep[modulo the different spatial resolutions of the observations]{Levy_etal_2021}.

Nonetheless, it is still interesting to compare the wind-driving capability of the YMCs to the multi-phase gas outflow seen at the moment.
Our Starburst99 simulation suggests that the YMCs can together produce a mechanical luminosity of ${\sim}2\times10^{40}\;\mathrm{erg\;s^{-1}}$, or a total deposited mechanical energy of ${\sim}3\times10^{54}\,$erg over 4~Myr.
A mere 2\% energy retention factor \citep[easily reachable for stellar wind and supernova-driven outflows; see][]{KimCG_etal_2020,Sirressi_etal_2024} would be enough to power the ionized gas outflow reported by \citet{Leaman_etal_2019} and to heat the X-ray emitting gas studied by \citet{Swartz_etal_2006}.
These order-of-magnitude calculations suggest that the current population of YMCs will be more than capable of drive gas outflow at a similar level in the future.
Therefore, the multi-phase gas outflow may get enhanced over the next $\sim$10~Myr and become closer to those seen in NGC~253 and NGC~4945 \citep[e.g.,][]{Westmoquette_etal_2011,Krieger_etal_2019,Bolatto_etal_2021}.


\section{Conclusions} \label{sec:conclusion}

In this paper, we examine a population of embedded YMCs in the central $1{\times}1$~kpc region of the nearby galaxy NGC~3351.
This system features a prominent central starburst ring fed by stellar bar-driven gas inflows from the outer disk \citep{Regan_etal_2006,Leaman_etal_2019}.
The proximity \citep[9.96~Mpc;][]{Anand_etal_2021}, favorable viewing angle \citep[$i\approx45^\circ$;][]{Lang_etal_2020}, and clean orbital configuration of this system (Figure~\ref{fig:wide_field} and \ref{fig:alma-zoom}) make it ideal for a multiwavelength YMC study in the full context of the large-scale host galaxy properties.

To this end, we acquire new ALMA data in Band~3 and 7 (project code 2021.1.00059.S; PI: J.~Sun), targeting the 93~GHz and 350~GHz continua along with various molecular lines.
The long-baseline observations ensure that the ALMA images reach similarly high resolution ($0\farcs1{-}0\farcs2$, equivalent to $5{-}10$~pc) as existing HST and JWST images \citep{Lee_etal_2022,Lee_etal_2023}, while the shorter-baseline observations guarantee robust imaging by recovering emission on larger spatial scales.
The jointly-imaged ALMA data are sensitive enough to detect thermal dust continuum and molecular line emission from the gas reservoir of individual forming YMCs, as well as free-free emission from \HII\ regions created by the YMCs.
By cross-matching ALMA sources with those in JWST and HST data, we probe the (often embedded) YMC stellar population, thereby achieving a complete, multiwavelength view of YMC formation.

Our joint analyses of the ALMA, HST, and JWST datasets yield the following key results:

\begin{enumerate}[leftmargin=1.2em,topsep=0.5em,itemsep=0.3em]

\item We find 18 bright, compact sources in the ALMA continuum images, with 15 detected at 93~GHz and 11 detected at 350~GHz (Figure~\ref{fig:alma-zoom}). Subsequent source cross-matching shows that only 8 of them have potential counterparts in JWST images (Figure~\ref{fig:jwst-zoom}) and 6 have counterparts in HST images (Figure~\ref{fig:hst-zoom}; also see Table~\ref{tab:crossmatch}).

\item Based on the ALMA continuum and molecular line measurements (Table~\ref{tab:YMC_obs}), we estimate for all sources their half-light radii ($1{-}8$~pc), stellar masses ($0.2{-}1{\times}10^5\,\uM$), gas masses ($0.3{-}2{\times}10^5\,\uM$), and gas velocity dispersion ($8{-}16\;\uV$). These estimates are comparable to typical values found for YMCs in the Milky Way and in other systems.

\item The estimated size, mass, and velocity dispersion also imply high total mass (${\gtrsim}10^5\,\uM$), a wide range of gas fractions (from ${\lesssim}25\%$ to ${\gtrsim}70\%$), large escape velocity ($6{-}10\;\uV$), short free-fall time ($0.5{-}1\;$Myr), and high gas surface density ($500{-}2000\;\uSig$), as summarized in Table~\ref{tab:YMC_phys}.
The last three quantities suggest that various forms of feedback (photoionization, supernova, and radiation pressure) may be less effective in regulating star formation for sources with such extreme conditions (see Sections~\ref{sec:results:Vesc}--\ref{sec:results:Sigmagas}).

\item The multiwavelength properties of these ALMA-identified sources motivate a classification scheme in which most of them belong to one of the following categories: starless clumps ($N=3$), clump--\HII\ region complexes (3), clump--\HII\ region--cluster complexes (5), and exposed \HII\ region--cluster complexes (3). These four categories likely represent four phases of YMC formation. The remaining 4 sources (detected only by ALMA at 93~GHz) are possibly supernova remnants with strong synchrotron emission rather than forming YMCs, though follow-up observations are necessary to verify this interpretation.

\item Comparing the number counts of ALMA-identified YMCs versus young HST clusters in a matched mass range, we infer an evolutionary timeline for forming YMCs (Figure~\ref{fig:timeline}). Modulo various sources of uncertainty, we estimate a duration of ${\sim}1{-}2$~Myr (or ${\sim}2$ free-fall times) for the starless clump phase. It then takes ${\sim}1{-}2$~Myr and ${\sim}2{-}3$~Myr for the newly formed cluster to become visible in the IR and optical bands, respectively. The cold gas reservoir disappears over ${\sim}3{-}4$~Myr (or $4{-}6$ free-fall times), and the \HII\ region disappears over ${\sim}4{-}6$~Myr. These numbers represent our best constraints on YMC formation timeline in an extragalactic, starburst-like environment. They also agree quantitatively with previous estimates by observational and numerical studies of various types of cluster-forming environments.

\item Putting the YMCs in the context of the entire central starburst region of NGC~3351, we find that the YMCs alone can account for at least 30--50\% of the total SFR of the ring ($0.3{-}0.5\,\uSFR$). While the YMCs exhibit an uneven azimuthal distribution and concentrate towards the two ``contact points,'' there is no consistent azimuthal trend in the inferred YMC evolutionary stages or other properties, as one may naively expect if YMC formation is only triggered by colliding flows near the ``contact points.'' Last but not least, the estimated total mechanical luminosity of the YMCs is large enough to power the previously reported multi-phase gas outflow from this system.

\end{enumerate}

The quantitative measurements presented in this study will likely be improved in follow-up studies based on existing and/or future observations.
For example, we plan to revisit the cluster selection and SED fitting by jointly analyzing the HST and JWST broad- and medium-band data shown in this work as well as archival HST narrow-band H$\alpha$ data \citep[PI: R.~Chandar; also see][]{Calzetti_etal_2021}.
The inclusion of deeper 93~GHz continuum observations with ALMA (partly executed in Cycle~9), lower frequency observations with VLA, or narrow-band imaging targeting IR recombination lines with JWST will certainly provide much better constraints on stellar age and extinction, eliminating a major source of uncertainty.

Beyond the YMCs, we also expect our rich ALMA dataset to support many other science goals.
The deep, highly resolved \CO32\ image makes it possible to characterize the complex, multi-scale gas structures present throughout NGC~3351's central region.
Detailed analysis of the gas kinematics may also shed light on the baryonic cycle across this system, from the large-scale gas inflow, to gas depletion and expulsion due to star formation and feedback, to the feeding (or lack thereof) of the central supermassive black hole.


\vspace{\baselineskip}
{
We appreciate helpful discussions with Blake~Ledger, Mark~Krumholz, Sumit~Sarbadhicary, and Todd~Thompson.

JS acknowledges support by the National Aeronautics and Space Administration (NASA) through the NASA Hubble Fellowship grant HST-HF2-51544 awarded by the Space Telescope Science Institute (STScI), which is operated by the Association of Universities for Research in Astronomy, Inc., under contract NAS~5-26555. JS also acknowledges support by the Natural Sciences and Engineering Research Council of Canada (NSERC) [funding reference number 568580] through a CITA National Fellowship.
The research of HH and CDW is supported by grants from NSERC and the Canada Research Chairs program.
The research of KB is supported by a Mitacs Globalink Research Award.
RCL acknowledges support for this work provided by the National Science Foundation (NSF) Astronomy and Astrophysics Postdoctoral Fellowship under award AST-2102625.
The work of AKL is partially supported by the NSF under Grants No.~1615105, 1615109, and 1653300. 
ECO acknowledges support from grant 510940 from the Simons Foundation. 
MC gratefully acknowledges funding from the DFG through an Emmy Noether Research Group (grant number CH2137/1-1). COOL Research DAO is a Decentralized Autonomous Organization supporting research in astrophysics aimed at uncovering our cosmic origins.
KG is supported by the Australian Research Council through the Discovery Early Career Researcher Award Fellowship (project number DE220100766) funded by the Australian Government. KG is supported by the Australian Research Council Centre of Excellence for All Sky Astrophysics in 3 Dimensions (ASTRO~3D), through project number CE170100013. 
JDH gratefully acknowledges financial support from the Royal Society (University Research Fellowship; URF/R1/221620).
MJJD and MQ acknowledge support from the Spanish grant PID2022-138560NB-I00, funded by MCIN/AEI/10.13039/501100011033/FEDER, EU.
RSK acknowledges funding from the European Research Council in the ERC Synergy Grant ECOGAL (grant 855130), from the German Excellence Strategy in the Heidelberg Cluster of Excellence STRUCTURES (EXC-2181/1 - 390900948), and from the German Ministry for Economic Affairs and Climate Action in project MAINN (funding ID 50OO2206).
MCS acknowledges financial support of the Royal Society (URF\textbackslash R1\textbackslash 221118).

This paper makes use of the following ALMA data: \linebreak
ADS/JAO.ALMA\#2013.1.00634.S, \linebreak 
ADS/JAO.ALMA\#2015.1.00956.S, \linebreak 
ADS/JAO.ALMA\#2021.1.00059.S, \linebreak 
ADS/JAO.ALMA\#2022.1.00159.S. \linebreak 
ALMA is a partnership of ESO (representing its member states), NSF (USA), and NINS (Japan), together with NRC (Canada), MOST and ASIAA (Taiwan), and KASI (Republic of Korea), in cooperation with the Republic of Chile. The Joint ALMA Observatory is operated by ESO, AUI/NRAO, and NAOJ. The National Radio Astronomy Observatory is a facility of NSF operated under cooperative agreement by Associated Universities, Inc.

This paper uses observations made with the NASA/ESA Hubble Space Telescope (HST) and the NASA/ESA/CSA James Webb Space Telescope (JWST). The data were obtained from the Mikulski Archive for Space Telescopes at the STScI, which is operated by the Association of Universities for Research in Astronomy, Inc., under NASA contract NAS~5-26555 for HST and NAS~5-03127 for JWST. The JWST observations are associated with program 2017.
All the HST and JWST data used in this paper are available at MAST:
\dataset[10.17909/t9-r08f-dq31]{http://dx.doi.org/10.17909/t9-r08f-dq31},
\dataset[10.17909/jray-9798]{http://dx.doi.org/10.17909/jray-9798},
\dataset[10.17909/ew88-jt15]{http://dx.doi.org/10.17909/ew88-jt15}.

We acknowledge the usage of the SAO/NASA Astrophysics Data System.

\facilities{
ALMA, HST, JWST
}

\software{
\texttt{NumPy} \citep{NumPy_2020},
\texttt{SciPy} \citep{SciPy_2020},
\texttt{Matplotlib} \citep{Matplotlib_2007},
\texttt{Astropy} \citep{Astropy_2013,Astropy_2018},
\texttt{CASA} \citep{CASA_2022},
\texttt{CARTA} \citep{CARTA_2.0.0},
\texttt{APLpy} \citep{APLpy_2012},
\texttt{adstex} (\url{https://github.com/yymao/adstex}).
}

}


\appendix

\section{Molecular Line Spectra for the ALMA Continuum Sources}
\label{apdx:spectra}

\begin{figure*}[p]
\centering
\gridline{
\fig{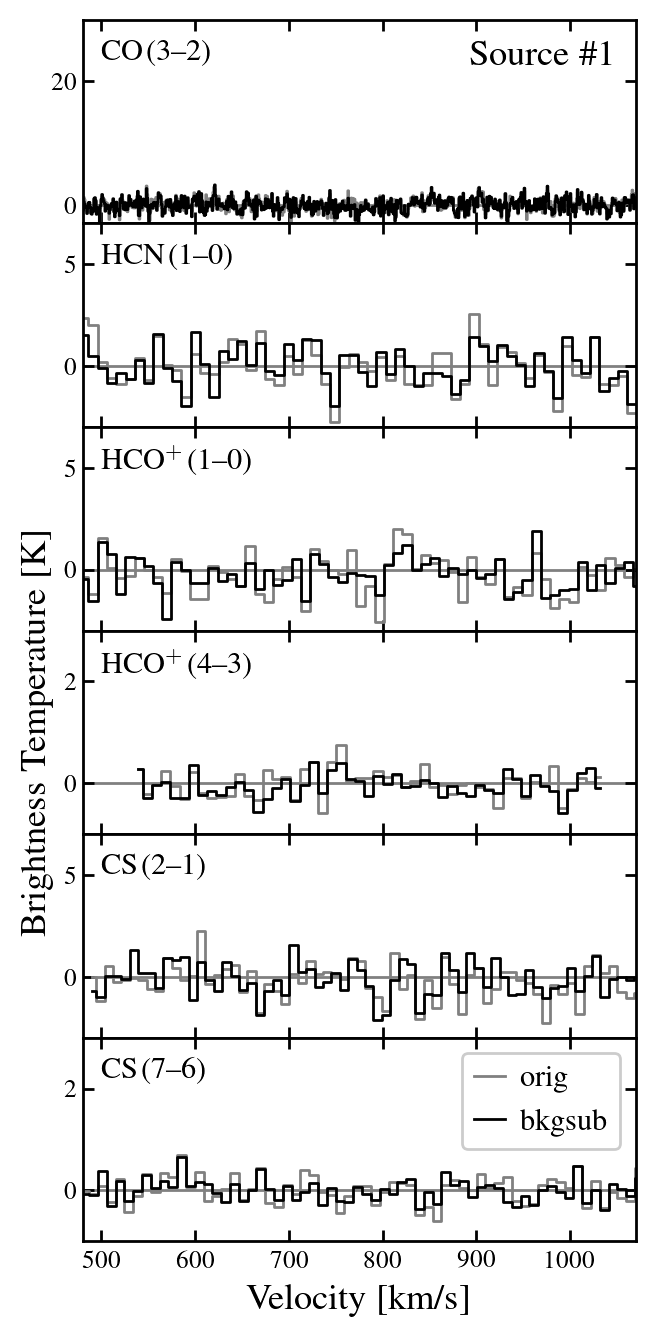}{0.49\textwidth}{}
\hfill
\fig{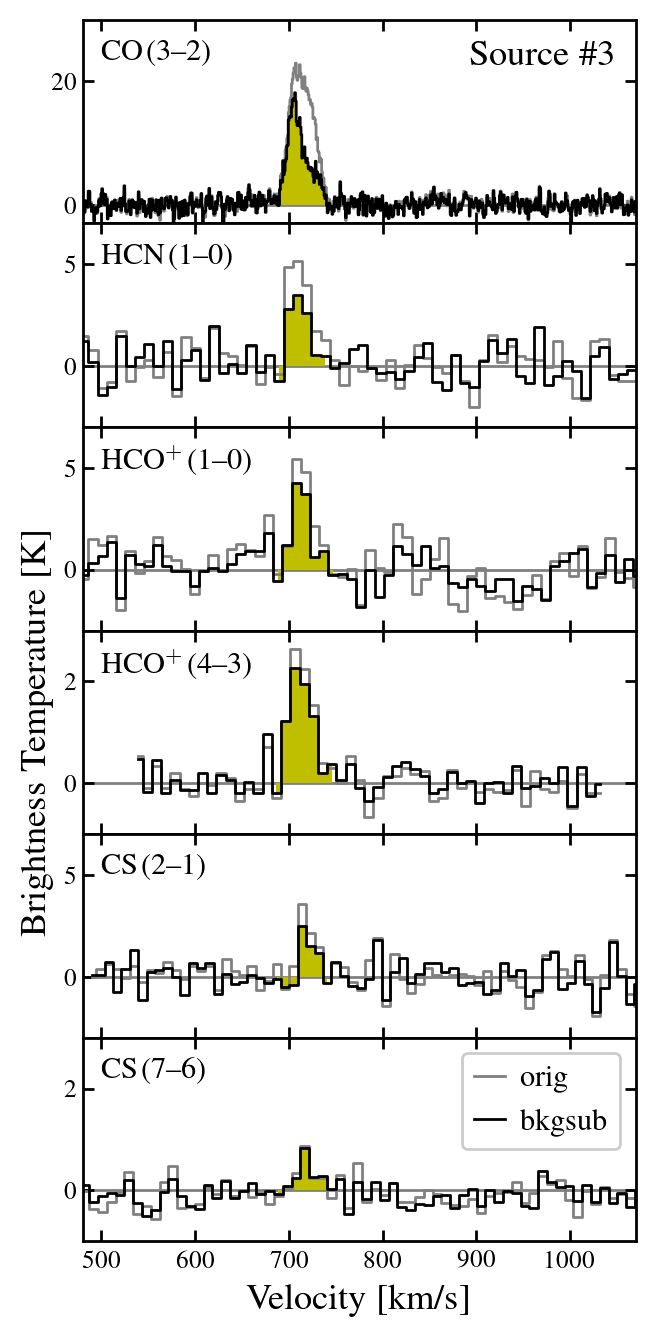}{0.49\textwidth}{}
}
\vspace{-2.0\baselineskip}
\caption{
Extracted molecular line spectra for two ALMA sources (\#1 and \#3).
The former represents a non-detection across all lines while the latter is detected in all lines at the same velocity (yellow shaded area).
In each panel, a grey line shows the original spectra extracted within the aperture defined for each source, whereas a black line shows the background-subtracted spectra (see Section~\ref{sec:analyses:characterize}).
The spectra for all 18 sources are available online as a figure set.
}
\vspace{0.5\baselineskip}
\label{fig:spectra}
\end{figure*}

We show the molecular line spectra for the ALMA continuum sources in Figure~\ref{fig:spectra}.


\bibliography{M95_YMC}




\suppressAffiliationsfalse
\allauthors


\end{document}